\documentclass[conference]{IEEEtran}
\usepackage{cite}
\usepackage{bbm}
\usepackage{siunitx}
\usepackage{epsfig}
\usepackage{textcomp}
\usepackage{amsfonts,amsmath,color,amssymb,amsxtra,graphicx}
\usepackage{amsmath}
\usepackage{amssymb}
\usepackage{bm}
\usepackage{subfigure}
\usepackage{graphicx}
\usepackage{algorithm}
\usepackage{algpseudocode}
\usepackage[skip=8pt,font=scriptsize]{caption}
\usepackage{pbox}
\usepackage[dvipsnames]{xcolor}

\usepackage[letterpaper, left=1in, right=1in, bottom=1in, top=0.75in]{geometry}
\usepackage{stackengine}
\def\delequal{\mathrel{\ensurestackMath{\stackon[1pt]{=}{\scriptstyle\Delta}}}}
\IEEEoverridecommandlockouts

\begin{document}
	
\title{Average Age-of-Information with a \\Backup Information Source}

\author{Elvina~Gindullina$^\dagger$, ~Leonardo Badia$^\dagger$ and ~Deniz G\"{u}nd\"{u}z$^\ddagger$ \\
	$^\dagger$ University of Padova, Dept. of Information Engineering, via Gradenigo 6B, 35131 Padova, Italy\\
	$^\ddagger$ Imperial College London, Dept. of Electrical and Electronic Engineering, London SW7 2AZ, U.K.\\  
	email: \{elvina.gindullina, leonardo.badia\}@dei.unipd.it, d.gunduz@imperial.ac.uk
\thanks{This work has received funding from the European Union's Horizon 2020 research and innovation programme under the Marie Sklodowska-Curie grant agreement No. 675891 \mbox{(SCAVENGE)}. D. G\"{u}nd\"{u}z also received funding from the European Research Council (ERC) through project BEACON (grant No. 725731)}
}


\maketitle

\begin{abstract}

Data collected and transmitted by Internet of things (IoT) devices are typically used for control and monitoring purposes; and hence, their timely delivery is of utmost importance for the underlying applications. However, IoT devices operate with very limited energy sources, severely reducing their ability for timely collection and processing of status updates.  IoT systems make up for these limitations by employing multiple low-power low-complexity devices that can monitor the same signal, possibly with different quality observations and different energy costs, to create diversity against the limitations of individual nodes. We investigate policies to minimize the average age of information (AoI) in a monitoring system that collects data from two sources of information denoted as \emph{primary} and \emph{backup} sources, respectively. We assume that each source offers a different trade-off between the AoI and the energy cost. The monitoring node is equipped with a finite size battery and harvests ambient energy. For this setup, we formulate the scheduling of status updates from the two sources as a Markov decision process (MDP), and obtain a policy that decides on the optimal action to take (i.e., which source to query or remain idle) depending on the current energy level and AoI. The performance of the obtained policy is compared with an aggressive policy for different system parameters. We identify few types of optimal solution structures and discuss the benefits of having a backup source of information in the system.

\end{abstract}


\section{Introduction}

Internet of things (IoT) is increasingly being deployed for critical operations such as factory and process automation, intelligent transportation and smart cities \cite{schulz2017latency}. Differently from other networks that are generally characterized in terms of throughput and delay, a key performance indicator for such applications is the \emph{age of information} (AoI), which quantifies the freshness of the destination's knowledge about the status of the system being monitored \cite{kaul2012real, kaul2011minimizing}. Another distinct aspect of these systems is that IoT devices are typically limited in available energy. One way to increase the sustainability of such a system is to exploit energy harvesting capabilities from ambient sources \cite{basagni2013wireless}. However, harvested energy is characterized by irregular energy arrivals randomly distributed over time; requiring a rechargeable battery as an energy buffer, and some form of intelligent control to avoid outages at critical instances \cite{gunduz2014designing}. In particular, if AoI is to be taken into account, status updates must be acquired sparingly depending on the level of energy available in the battery.

A growing number of papers investigate the evolution and control of AoI in energy-harvesting systems  \cite{bacinoglu2017scheduling, arafa2018age, arafa2018agemin, feng2018age, ceran2019reinforcement}. The scenario of reference involves a device making optimal decisions about acquiring status updates depending on the energy cost and the available battery level.
In this paper, we consider instead an IoT system that can exploit \emph{multiple} sources of information, each providing a different energy-age trade-off. For example, an IoT device may exploit multiple sensors with different reliabilities and costs. Alternatively, we may think of a terminal that can update the system status through either a cellular technology, which guarantees reliability and high coverage, but is very expensive in terms of energy, or a low-range energy-aware technology.

For the purpose of the analysis we refer to the following model. We consider two \emph{sources} providing information with different costs and qualities (freshness and/or reliability) to a \emph{monitoring node} that tries to optimize the resulting AoI over time within a constrained energy budget. The information source with higher cost and quality is called the \emph{primary} source of information, while the other is referred to as the \emph{backup} source. These sources provide the monitoring node with the most fresh status update in their buffer. Therefore, the monitoring node does not know  with certainty the AoI of the packet that will be delivered from a source node. The only assumption the monitoring node can make is the reliability of a source, i.e., the probability of receiving a fresh data packet from that source. 
 
Minimization of AoI in a multi-source system is also considered in \cite{tripathi2017age}, where multiple sensors communicate with the monitoring system via orthogonal channels. However, in \cite{tripathi2017age} each sensor monitors a different process, and the objective is a function of the ages of all the processes. The proposed policy converts the scheduling problem into a bipartite matching problem  between the sets of channels and sensors. Similarly, in \cite{yates2018age}, \cite{pappas2015age} multiple sources provide status updates to a monitoring node about different processes, which results in a multi-objective problem, and the goal is to schedule transitions in order to balance the AoI of these different processes. In contrast with the aforementioned contributions, we consider a system with energy harvesting capabilities, where the monitoring node selectively requests status updates of the same underlying process from different nodes. We justify having more than one information source in the system to obtain diversity by balancing across different combinations of energy cost and AoI. This contribution can be considered as a first step towards developing policies for a multi-source IoT-system that can achieve efficient management without any prior knowledge of the cost or quality of the information sources, which have to be learned over time.

The rest of this paper is organized as follows. In Section \ref{sec:mod}, the system model description, problem formulation and solution approaches are introduced. Numerical results are presented in Section \ref{sec:res}, providing a performance comparison between the proposed solution and a reference approach, namely, the aggressive policy. The paper is concluded in Section \ref{sec:con}, where possible further developments are also outlined.

\section{System overview}
\label{sec:mod}

We consider a system consisting of a single energy-harvesting monitoring node and two sources of information, where each source takes measurements of the same underlying process that is of interest to the monitoring node. Time is discretized into time slots with a unit slot length of arbitrary duration. At each time slot, the monitoring node can receive a status update from only one of the sources. The status update from the chosen source becomes available to the monitoring node at the beginning of the time slot. 

The monitoring node consumes different amounts of energy to receive a status update from the two sources. We assume that the status updates provided by the two sensors are of age either $\alpha$ or $\beta$, referred to as \emph{fresh} and \emph{stale}, respectively, with $\alpha{<} \beta$. For the sake simplicity, we consider only two possible age values $\alpha$ and $\beta$, in this paper, which can model, for example, useful and useless data packets.
We assume that source $i$ can provide a fresh status update at each time slot with probability $\gamma_i$, $i=1,2$, and a stale packet with probability $1 - \gamma_i$, such that $\gamma_1 {>} \gamma_2$ for \emph{primary} and \emph{backup} sources, respectively. The AoI at the monitoring node increases by 1 if no new update is received.

The energy costs of requesting a status update from source $i$ is denoted by $c_i$, $i = 1,2$, where we assume $c_1 {>} c_2$. Here $c_1, c_2 \in \mathbb{Z}^+$ correspond to integer multiples of a unit of energy.

Battery level $b(t)$ is updated at each time slot depending on the energy harvested in the previous time slot and the energy cost of receiving a data packet from one of the sources: 
\vspace{-0.1cm}
\begin{equation}
\small
b(t)=\min\{b(t-1) - \sum_{i=1}^2 c_i \cdot \mathbbm{1}{(a(t) = a_i)} + e(t), B\},
\end{equation}
where $e(t) \in \{0, \bar{e}\}$ denotes the harvested energy available to be used in time slot $t$, $B$ is the battery capacity, and $\mathbbm{1}(x)$ is an indicator function: $\mathbbm{1}(x) = 1$ when $x$ holds, and $\mathbbm{1}(x) = 0$ otherwise. We assume $\{e(t)\}_{t=1}^\infty$ is an independent and identically distributed (i.i.d.) binary random process with $P(e(t) = \bar{e}) = \lambda$.

The monitoring node makes a decision at the beginning of each time slot whether to request a new status update or not, and if so, which source to request it from. We seek the policy that minimizes the average AoI at the monitoring node by optimally choosing the action to take at each time slot, accounting for the battery level and the current age of information. We first formulate the problem as an MDP. 

\subsection{MDP formulation}

An MDP consists of a tuple ${<}S, A, P, R{>}$ of state space $S$, action space $A$, probability transition function $P$, and a reward or cost function $R$. In our problem, finite space of actions $A$ includes requesting an update from either of the two  sources (primary/backup) and remaining idle. We set $A = \{a_0, a_1, a_2\}$, where $a_0$ corresponds to remaining idle, $a_1$ updating from the primary source, and $a_2$ updating from the backup source. 

Action $a_i$ is not allowed if $b(t) {<} c_i$. This can be incorporated into the framework with the same action space by imposing very high energy costs for action $a_i$ when $b(t) < c_i$, $i = 1, 2$. 

Let $\delta(t) \in \{1, 2, ..., \delta_{max}\}$ denote the AoI at the monitoring node at time slot $t$, where $\delta_{max}$ is the maximum age in the system. Equivalently, we assume that having a status information of age $\delta_{max}$, or any $\delta > \delta_{max}$ have the same utility. Depending on action $a(t)$, $\delta(t)$ can take one of the following values $\{\delta(t-1)+1, \alpha, \beta\}$ . The system state is described by the pair of variables $s(t) = (b(t), \delta(t))$. Note that we have a finite state space of dimension $(\delta_{max} + 1)(B+1)$. In this paper, we set $\beta$ as the maximum AoI, i.e., $\beta = \delta_{max}$, beyond which increase in age becomes irrelevant. Accordingly, receiving a stale status update is equivalent to not receiving a useful update. 

$P$ denotes the transition probabilities of the MDP, where $P(s'|s, a) = Pr(s(t+1) = s'|s(t) = s, a(t) = a)$; that is, the probability that taking action $a$ at state $s$ will lead to a transition to state $s'$ in the following time slot. The transition probabilities for our problem are given as follows for $a_i \in \{a_1, a_2\}$:
\vspace{-0.1cm}
\begin{equation}
	\small
	\begin{cases} 
	P((\min\{b {+} \bar{e} {-} c_{i}, B\}, \min\{\alpha, \delta{+}1\})|(b, \delta), a_i) = \lambda \gamma_i\\
		
	P((\min\{b {+} \bar{e} {-} c_{i}, B\}, \min\{\beta, \delta{+}1\})|(b, \delta), a_i) = \lambda (1{-}\gamma_i)\\
	
	P((b {-} c_{i}, \min\{\alpha, \delta{+}1\})|(b, \delta), a_i) = (1{-}\lambda) \gamma_i\\	
		
	P((b {-} c_{i}, \min\{\beta, \delta{+}1\})|(b, \delta), a_i) = (1{-}\lambda) (1{-}\gamma_i)\\	
				
	\end{cases}
\end{equation}

Note that, if the received status update is older than the currently available one, then the monitoring node drops the new packet and keeps the previous status update. We can conclude that if $\delta_t {<} \alpha$, then the optimal action is to remain idle, i.e., $a_t = a_0$.

When  the node remains idle, i.e., $a_t = a_0$, the transition probabilities are given as follows:
\vspace{-0.1cm}
\begin{equation}
\small
\begin{cases} 
P((b, \delta+1) | (b, \delta), a_0) = 1 - \lambda & \text{$b{<}B$} \\

P((\min\{b + \bar{e}, B\}, \delta+1) | (b, \delta), a_0) = \lambda & \text{$b{<}B$} \\

P((B, \delta+1) | (B, \delta), a_0) = 1 \\
\end{cases}
\end{equation}

The policy $\pi$ defines an action $a(t)$ at each time slot depending on the current state. The infinite-horizon time average AoI, when policy $\pi$ is employed, starting from initial state $s_0$, is defined as \cite{ceran2019average}:
\vspace{-0.2cm}
\begin{equation}
\small
V^{\pi}(s_0) = \lim \sup_{T \rightarrow \infty} \frac{1}{T} \mathbbm{E}\left[ \sum_{t=0}^T \delta^{\pi}(t) | s(0) = s_0
\right].
\end{equation}

A policy is optimal if it minimizes the average AoI - $V^{\pi}(s_0)$. The optimal infinite-horizon average AoI for a starting state $s_0$ is found by solving:
\vspace{-0.1cm}
\begin{equation}
\small
V(s_0) = \min_{\pi} V^{\pi}(s_0).
\end{equation}

To solve this optimization problem, we can use the offline dynamic programming approach adopting the relative value iteration (RVI) algorithm described in \cite{bertsekas1995dynamic}. In the offline approach we model the state transition function based on the statistical prior knowledge of the information sources' reliability and environmental characteristics. The RVI differs from VI by the value function of some state $V(s^*)$ in each update. In this case, the Bellman equation is defined as:
\vspace{-0.1cm}
\begin{equation}
\small
V^{n}(s) = \min_{a \in A} \Big(\delta(s, a) - V^{n-1}(s^*) + \sum_{s' \in S}P(s'|s,a)V^{n-1}(s')\Big),
\end{equation}
where $V^n$ is the value function, and $s^*$ is a fixed state chosen arbitrarily.

The optimal stationary deterministic policy, obtained by Algorithm \ref{VI}, specifies the decision rule that maps  the current energy level and AoI to actions taken with probability one. In Algorithm \ref{VI}, $sp(V^n - V^{n - 1}) {<} \epsilon$ stands for the stopping criteria, where $sp(V) = \max_{s \in S} V(s) - \min_{s \in S} V(s)$. We run the RVI algorithm until the stopping criteria holds. At that moment the policy $\pi$ achieves an average-cost AoI that is within $\epsilon \cdot 100\%$ of optimal.
\vspace{-0.1cm}
\begin{algorithm}
	\caption{Relative Value Iteration Algorithm}\label{VI}
	\small
	\begin{algorithmic}
		\State set $v^0(s) = 0$, $\forall s \in S$
		\State set n = 1, $\epsilon {>} 0$  
		\Repeat
			\State $n \leftarrow n+1$ 
			\ForAll {$s \in S$}
				\begin{equation*}
				\small
					\begin{split}
					\quad v^n(s) &= \min_{a \in A} \sum_{s' \in S}P(s'|s,a) \Big[\delta(s'|s,a) + V^{n-1}(s')\Big] \\
					 \quad V^n(s) &= v^n(s) - v^n(s_0)
					\end{split}
				\end{equation*}
				\text{$\quad$ where $s_0$ is a fixed state chosen arbitrary}
			\EndFor
		\Until {$sp(V^n - V^{n - 1}) {<} \epsilon$}
		\State \Return $arg \min V(s)$
	\end{algorithmic}
\end{algorithm} 

\section{Numerical results}
\label{sec:res}

In this section, we analyze the optimal policies for different settings, in particular, we consider the cost ratios between the primary and backup information sources, reliability of the sources, and the parameters of the energy harvesting process ($\lambda$, $\bar{e}$). We study the structure of the optimal policy, and try to identify the added value in average AoI from employing an extra information source in the system.

\subsection{Simulation parameters}

System parameters that remain constant for all the numerical simulations are presented in Table \ref{tab:parameters}. The efficiency of the optimal policy is verified via simulations run over $T = 5000$ time slots, and compared with a so-called \emph{aggressive policy}. The aggressive policy (Algorithm \ref{aggressive_policy}) tries to always receive a status update whenever it has sufficient energy in its battery, and goes for the expensive source whenever it can afford it.
\vspace{-0.1cm}
\begin{algorithm}
	\caption{Aggressive Policy}\label{aggressive_policy}
	\small
	\begin{algorithmic}
		\State set $b(0) = 0$, $\delta(0) = 0$ 
		\For{$t = 1:T$}
		{
			\If {$b(t) {\geq} c_1 $}
				\State $b(t) = \min\{b(t-1) + e(t) - c_1, B\}$
				\If {${p {\leq} \gamma_1}$} 
					\State $\delta(t) {=} \alpha$
					\Else
						\State $\delta(t) {=} \{\beta{:} \delta(t{-}1) {\geq} \beta; \delta(t{-}1){+}1{:} \delta(t{-}1) {<} \beta\}$
				\EndIf
			\ElsIf {$c_2 \leq b(t) {<} c_1$}
				\State $b(t) = \min\{b(t-1) + e(t) - c_2, B\}$
				\If {${p {\leq} \gamma_2}$} 
					\State $\delta(t) {=} \alpha$
				\Else
					\State $\delta(t) {=} \{\beta{:} \delta(t{-}1) {\geq} \beta; \delta(t{-}1){+}1{:} \delta(t{-}1) {<} \beta\}$
				\EndIf
			\Else
			\State $\delta(t) = \delta(t{-}1) {+} 1$, 
			\State $b(t) = \min\{b(t{-}1) {+} e(t), B\}$
			\EndIf}
		\EndFor
	\end{algorithmic}
\end{algorithm}

Denoting by $\bar{\delta}_T^m$ the time-average AoI over T time slots at the m-th run of the simulations, we consider the mean AoI $\bar{\delta}_T = \frac{1}{M} \sum_{m = 1}^M \bar{\delta}_T^m$ and its standard deviation:
\begin{equation}
\small
	st \delequal \sqrt{\frac{\sum_{m=1}^M (\bar{\delta}_T^m - \bar{\delta}_T)^2}{M-1}},
\end{equation} 
over $M = 1000$ runs of the simulations for each settings.

\begin{table}[t]
	\caption{Default parameters.}
	\label{tab:parameters}
	\centering 
	\begin{tabular}[h]{|p{6.0cm}|c|}
		\hline
		{\bf Parameters} & {\bf Values}\\
		\hline
		Battery capacity, $B$ & $20$ \\
		\hline
		Maximum age in the system, $\delta_{max}$ & $30$ \\
		\hline
		AoI states, $[\alpha, \beta]$ & $[1,20]$ \\
		\hline
		Amount of harvested energy per time slot, $\{0, \bar{e}\}$ & $\{0, 3\}$ \\
		\hline
		Reliability of the primary source, $\gamma_1$ & $0.9$ \\
		\hline
		\end{tabular}
\end{table}

\subsection{Cost ratio}

\begin{figure}[!t]%

	\subfigure[Cost ratio = 0.0, $\lambda = 0.2$, $\gamma_2 = 0.2$]{%
		\includegraphics[width=0.14\textwidth]{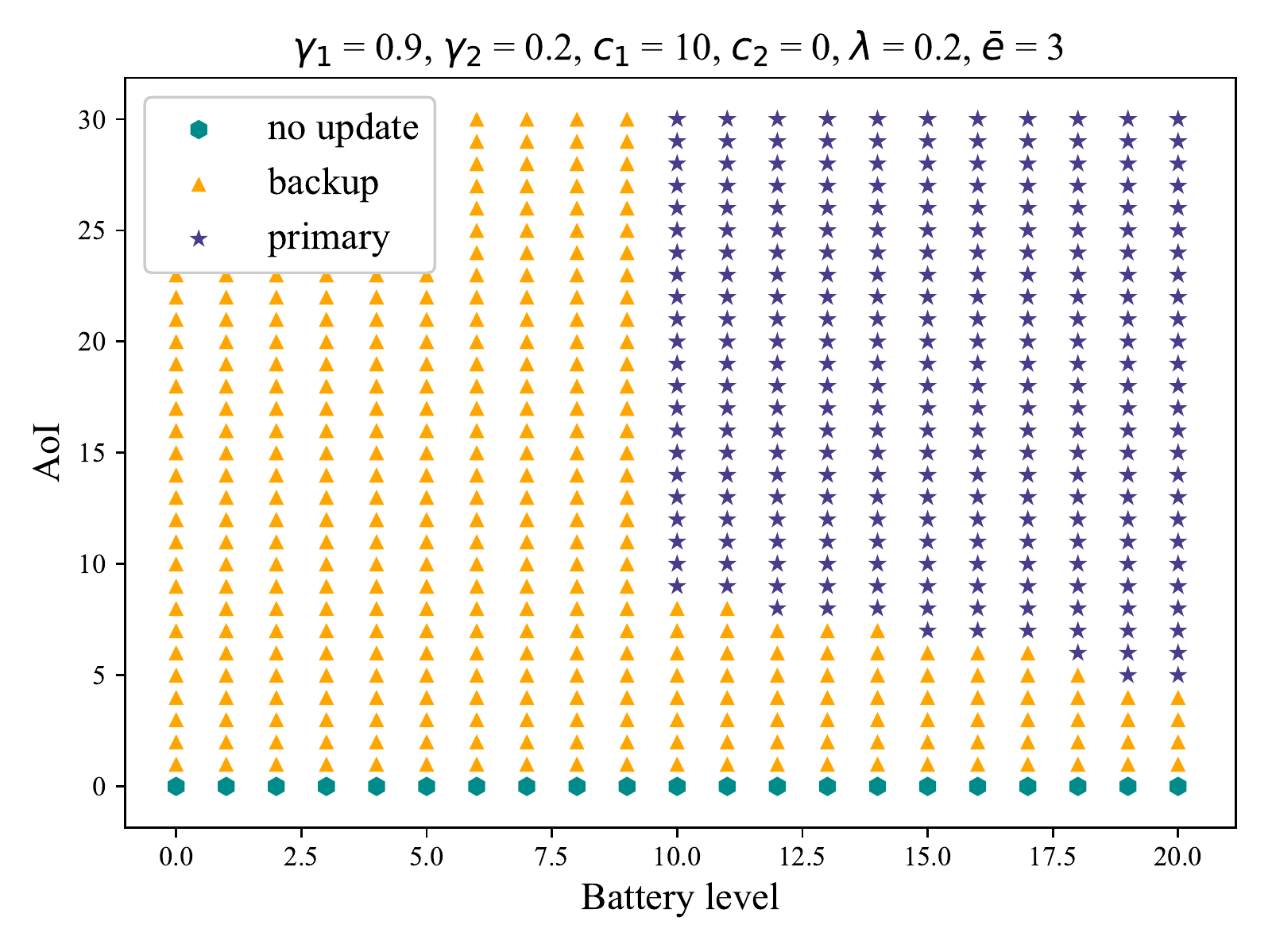}
		\label{fig:l02_g02_cr_00}%
	}
	\hspace*{\fill}
	\subfigure[Cost ratio = 0.4, $\lambda = 0.2$, $\gamma_2 = 0.2$]{%
		\includegraphics[width=0.14\textwidth]{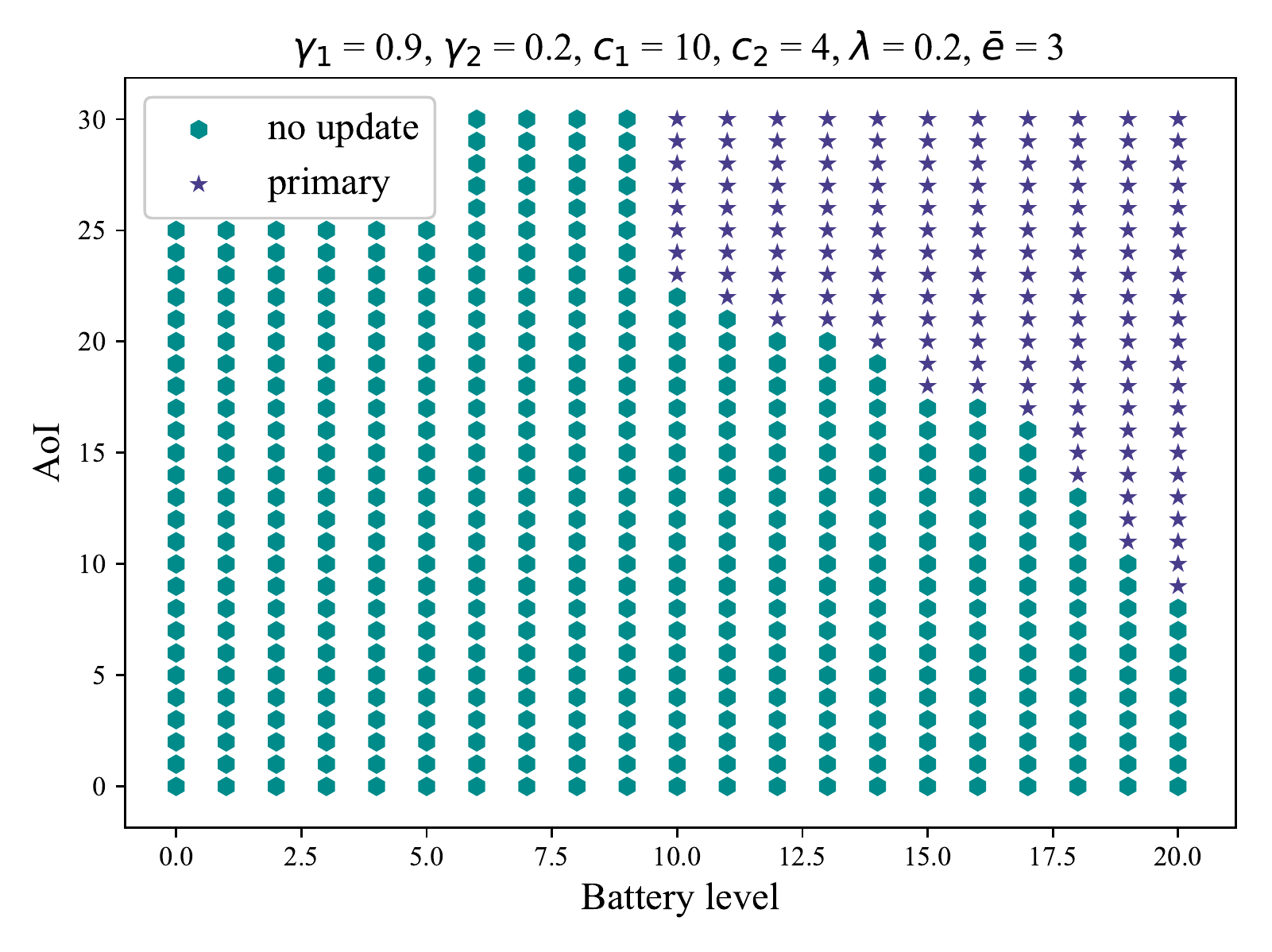}
		\label{fig:l02_g02_cr_04}%
	}
	\subfigure[Cost ratio = 0.8, $\lambda = 0.2$, $\gamma_2 = 0.2$]{%
		\includegraphics[width=0.14\textwidth]{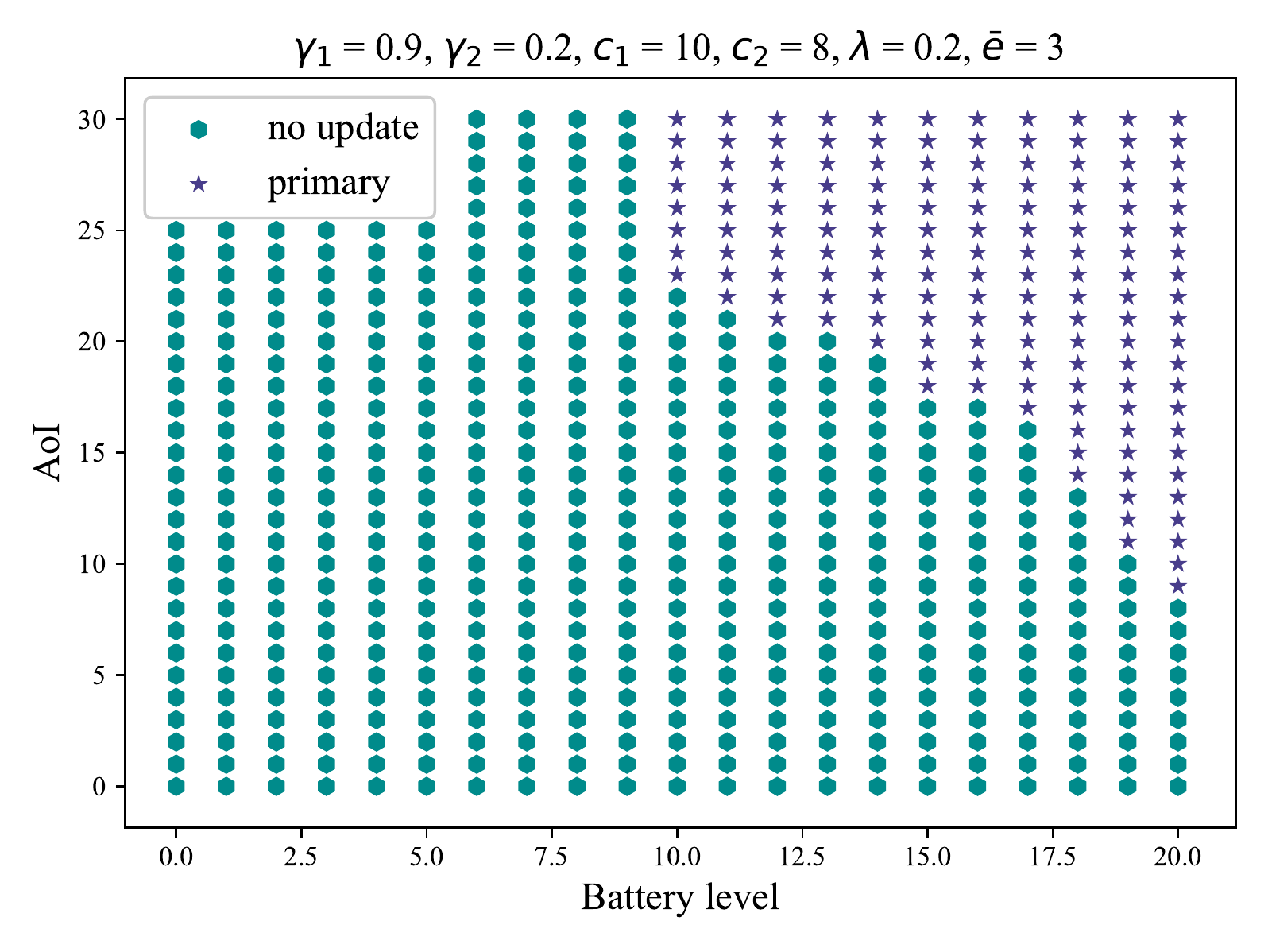}
		\label{fig:l02_g02_cr_08}%
	}
	\hspace*{\fill}
	\subfigure[Cost ratio = 0.0, $\lambda = 0.2$, $\gamma_2 = 0.8$]{%
		\includegraphics[width=0.14\textwidth]{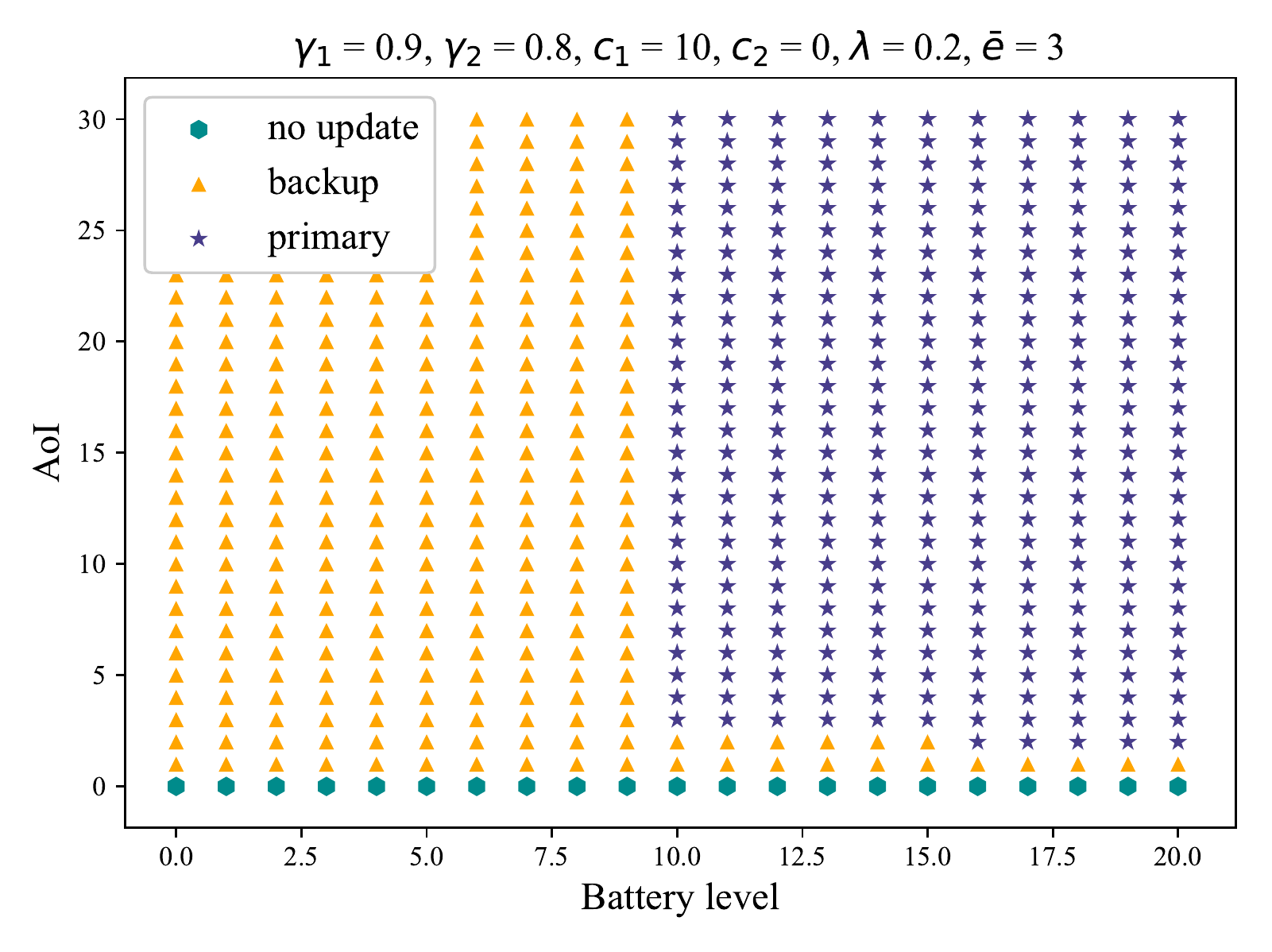}
		\label{fig:l02_g08_cr_00}%
	}
	\hspace*{\fill}
	\subfigure[Cost ratio = 0.4, $\lambda = 0.2$, $\gamma_2 = 0.8$]{%
		\includegraphics[width=0.14\textwidth]{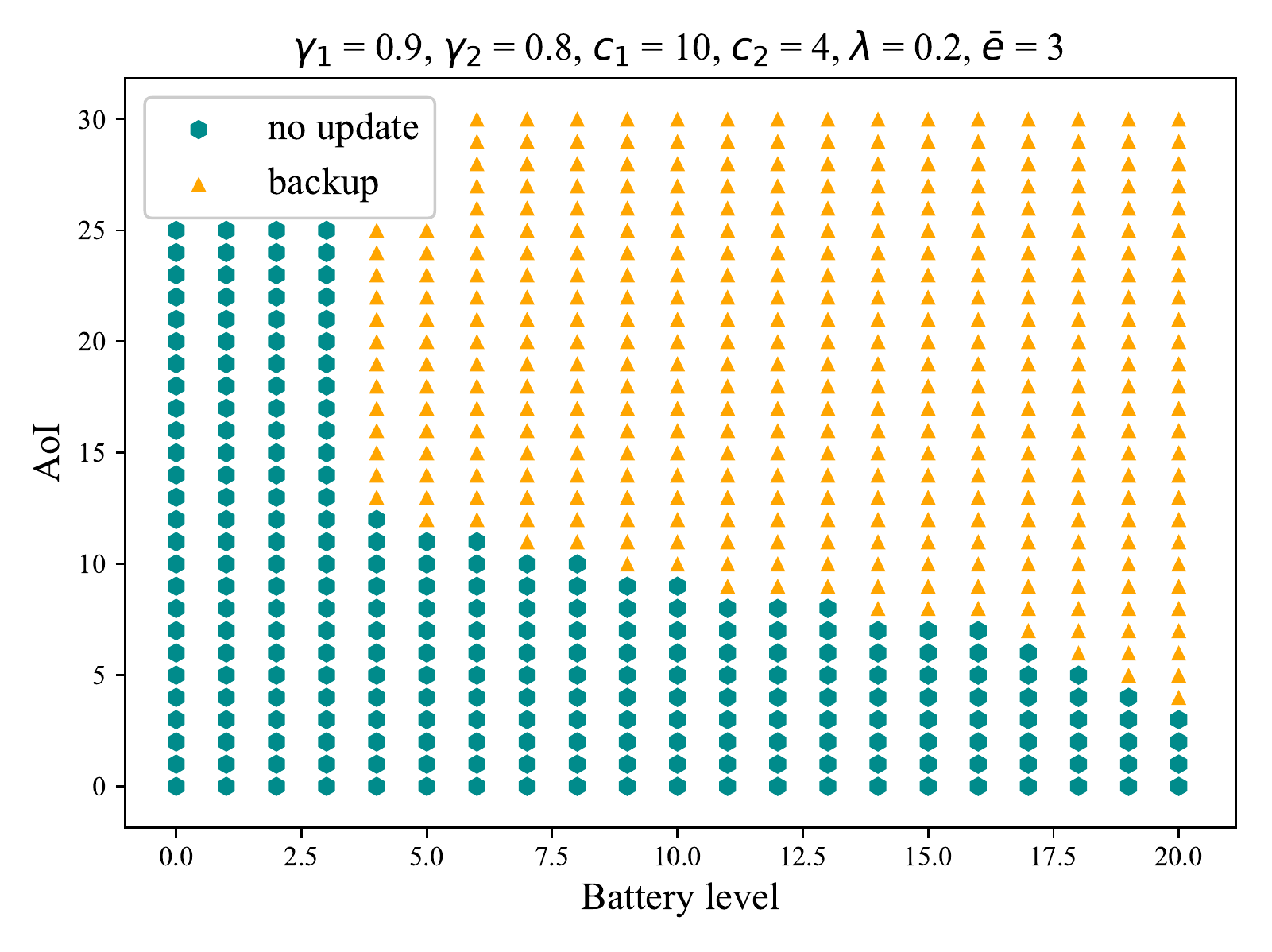}
		\label{fig:l02_g08_cr_04}%
	}
	\subfigure[Cost ratio = 0.8, $\lambda = 0.2$, $\gamma_2 = 0.8$]{%
		\includegraphics[width=0.14\textwidth]{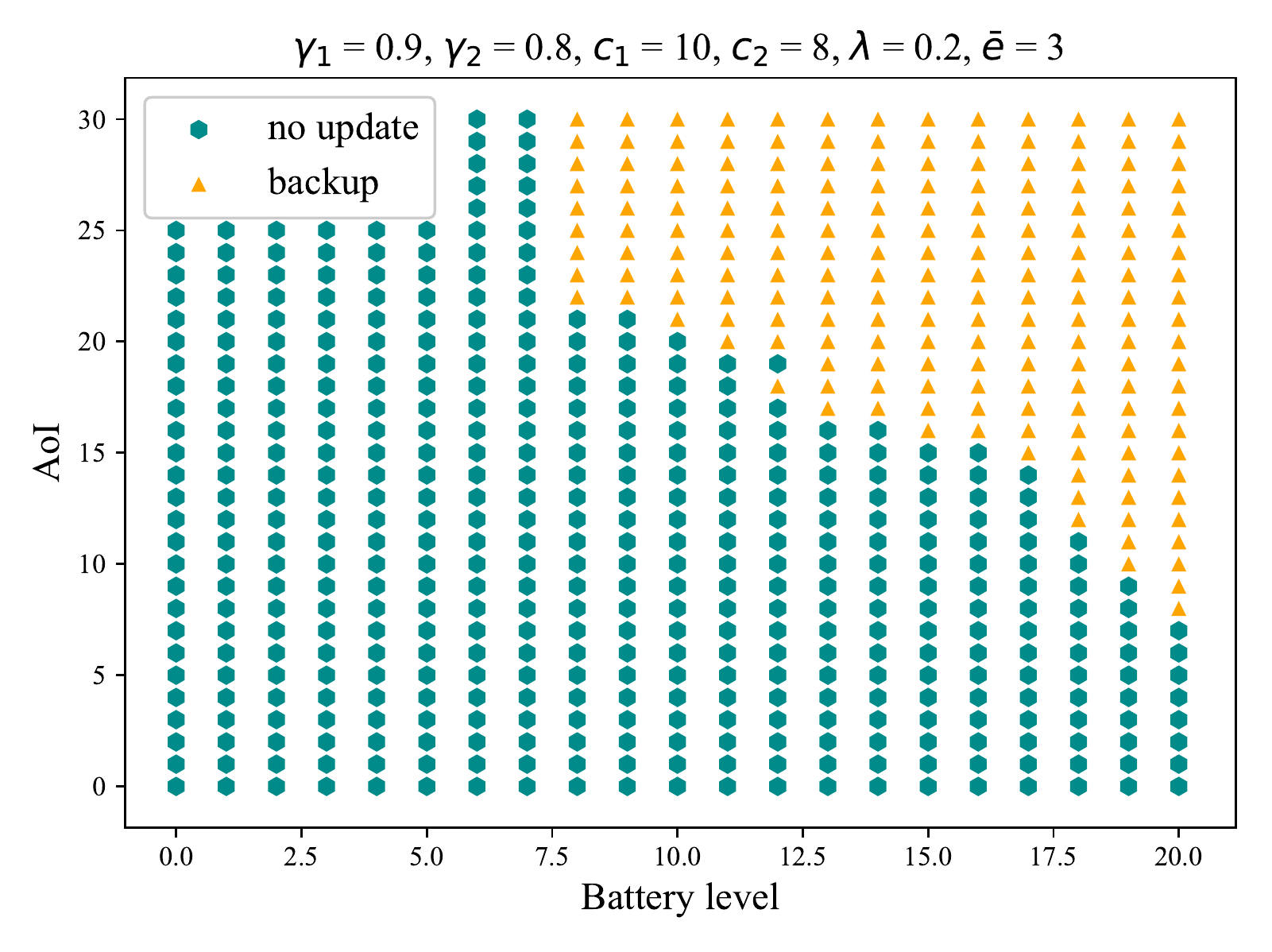}
		\label{fig:l02_g08_cr_08}%
	}
	\hspace*{\fill}
	
	\subfigure[Cost ratio = 0.0, $\lambda = 0.8$, $\gamma_2 = 0.2$]{%
		\includegraphics[width=0.14\textwidth]{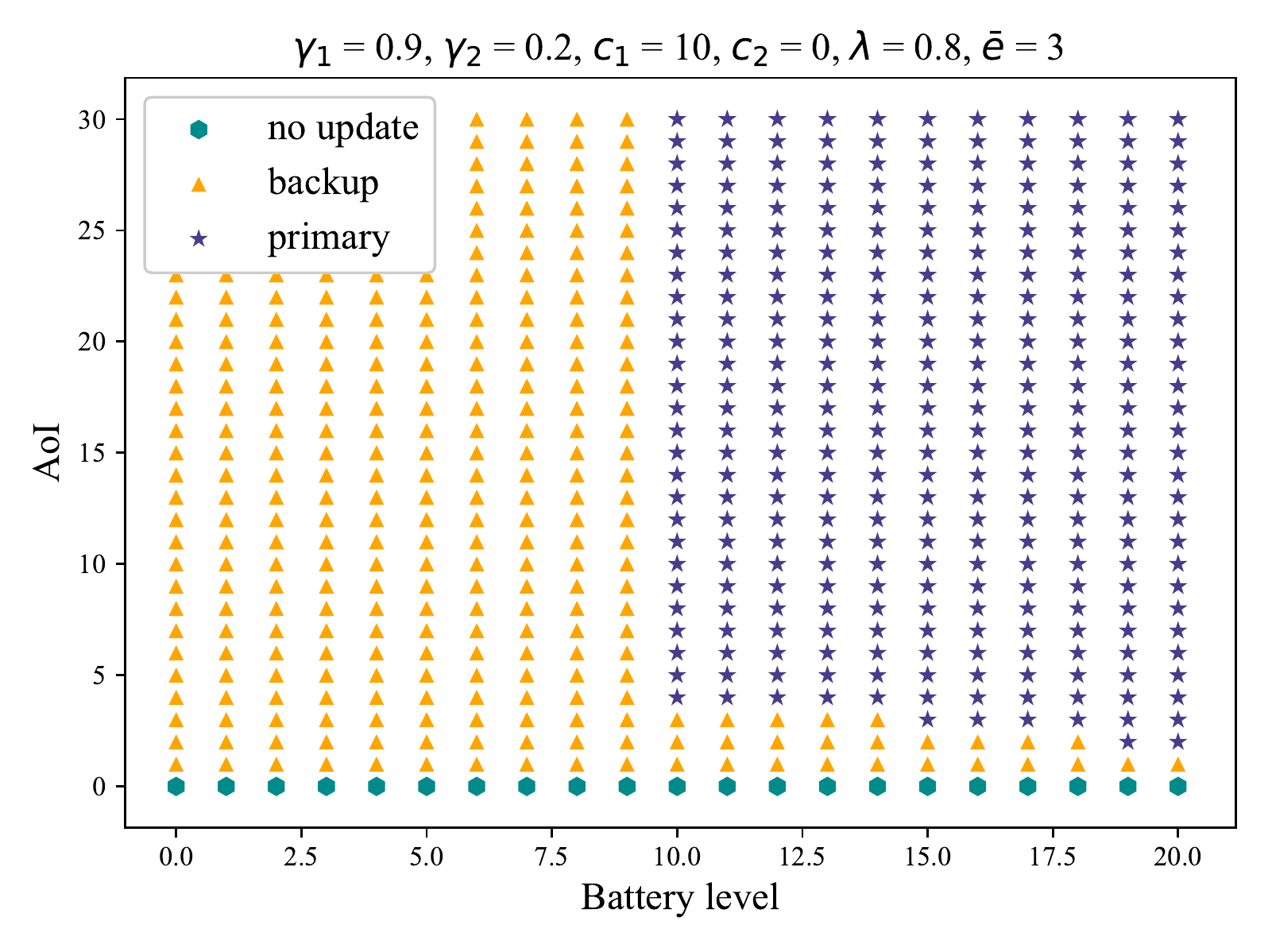}
		\label{fig:l08_g02_cr_00}%
	}
	\hspace*{\fill}
	\subfigure[Cost ratio = 0.4, $\lambda = 0.8$, $\gamma_2 = 0.2$]{%
		\includegraphics[width=0.14\textwidth]{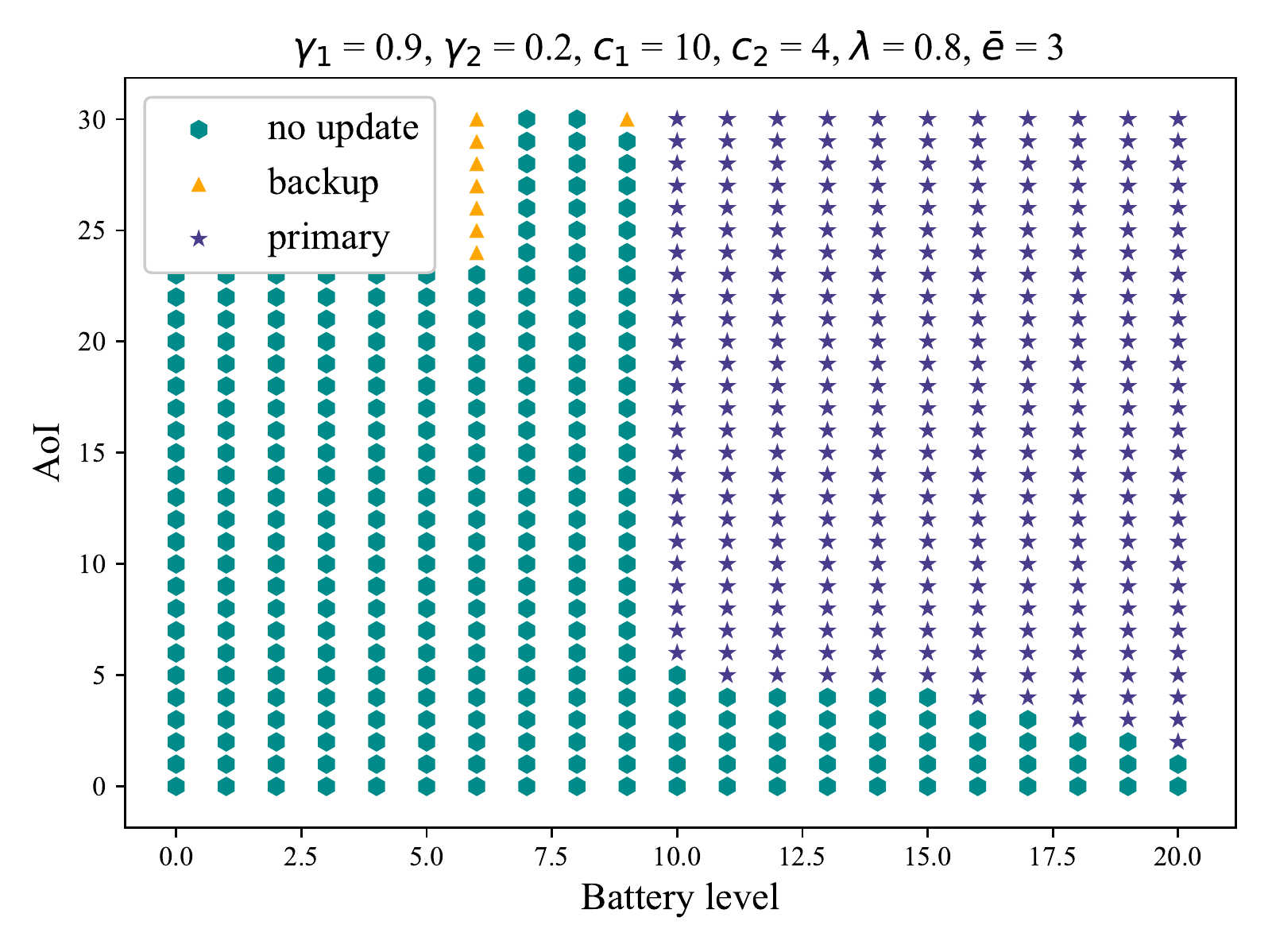}
		\label{fig:l08_g02_cr_04}%
	}
	\subfigure[Cost ratio = 0.8, $\lambda = 0.8$, $\gamma_2 = 0.2$]{%
		\includegraphics[width=0.14\textwidth]{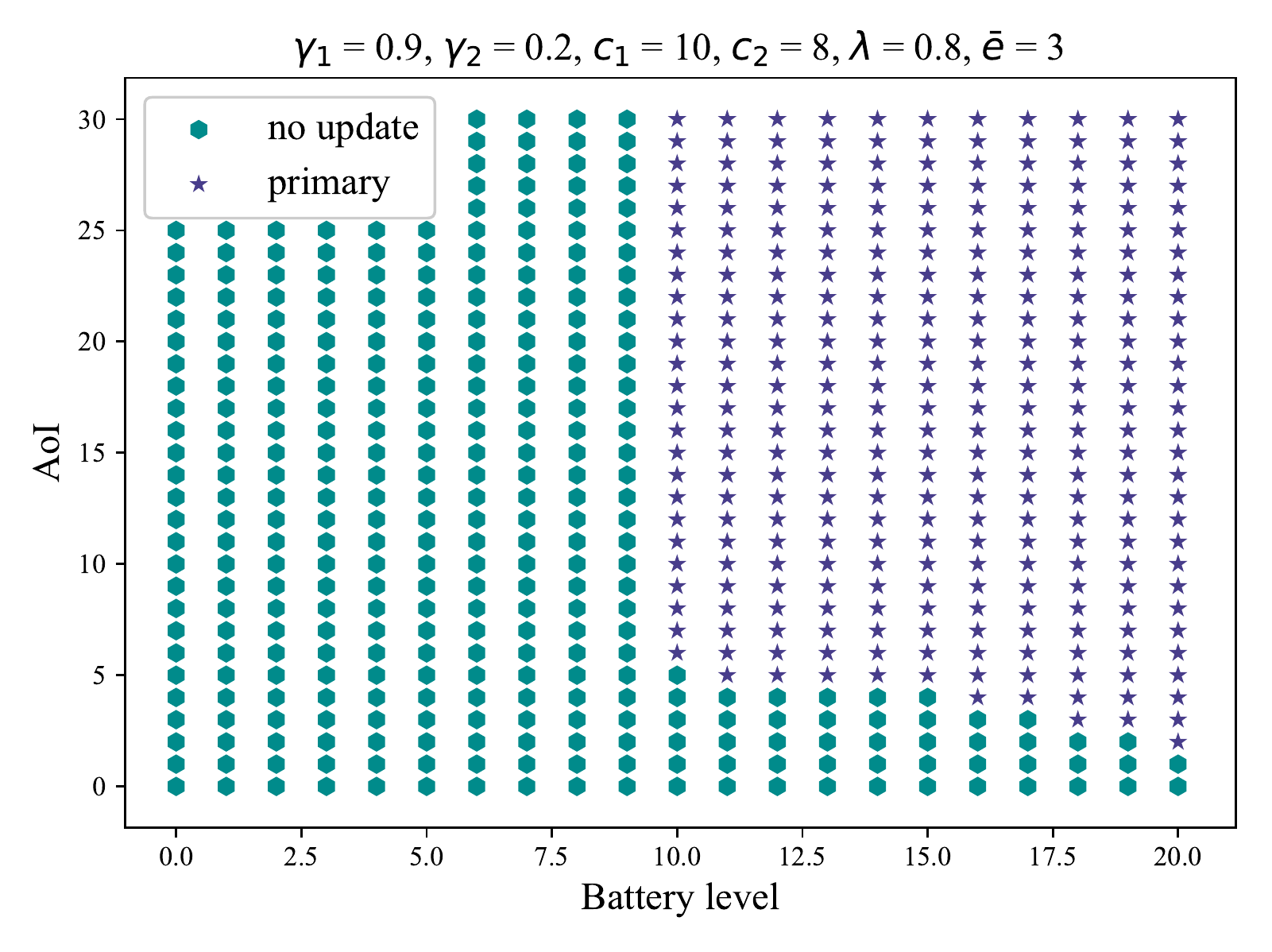}
		\label{fig:l08_g02_cr_08}%
	}
	\hspace*{\fill}
	\subfigure[Cost ratio = 0.0, $\lambda = 0.8$, $\gamma_2 = 0.8$]{%
		\includegraphics[width=0.14\textwidth]{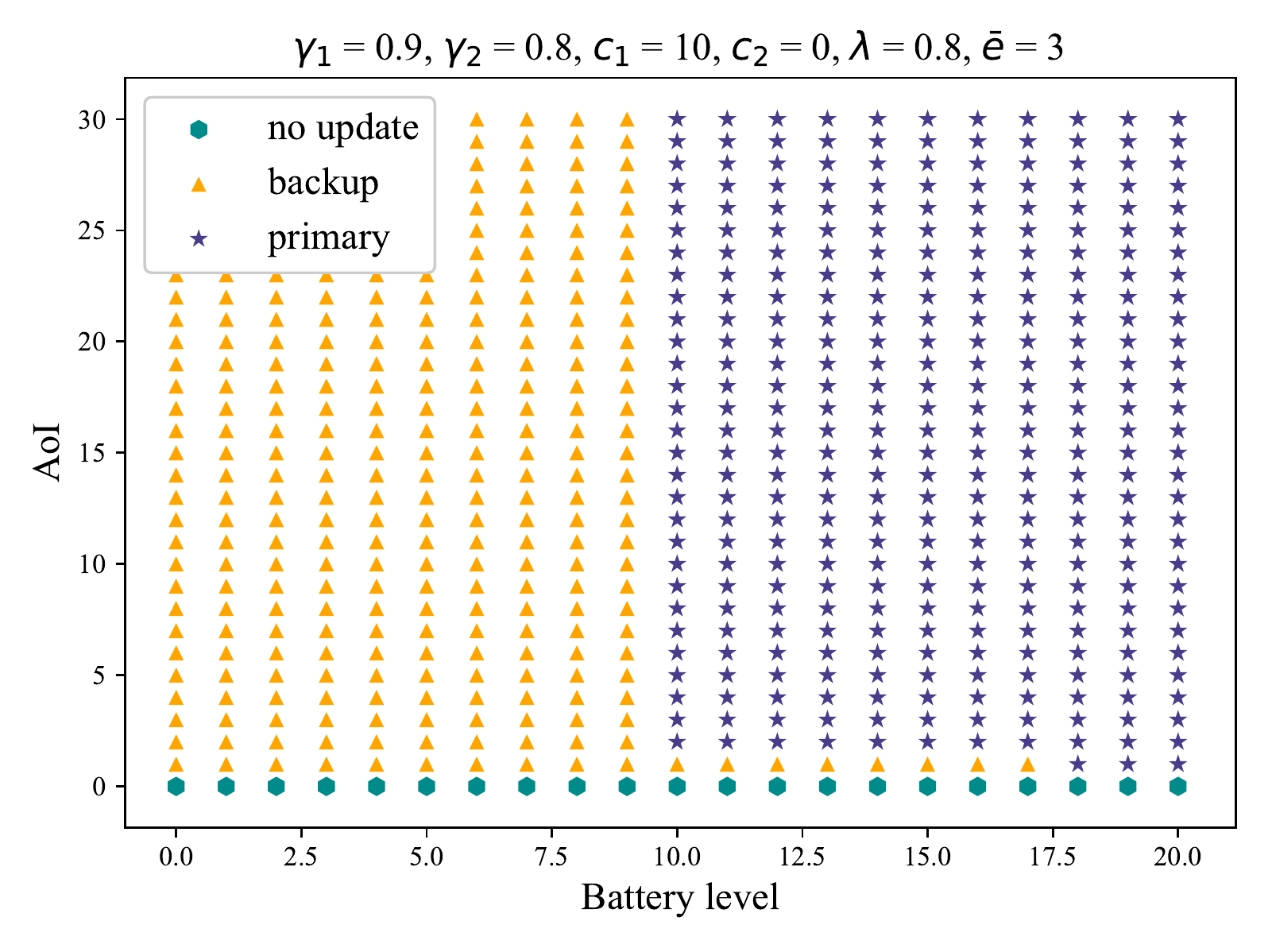}
		\label{fig:l08_g08_cr_00}%
	}
	\hspace*{\fill}
	\subfigure[Cost ratio = 0.4, $\lambda = 0.8$, $\gamma_2 = 0.8$]{%
		\includegraphics[width=0.14\textwidth]{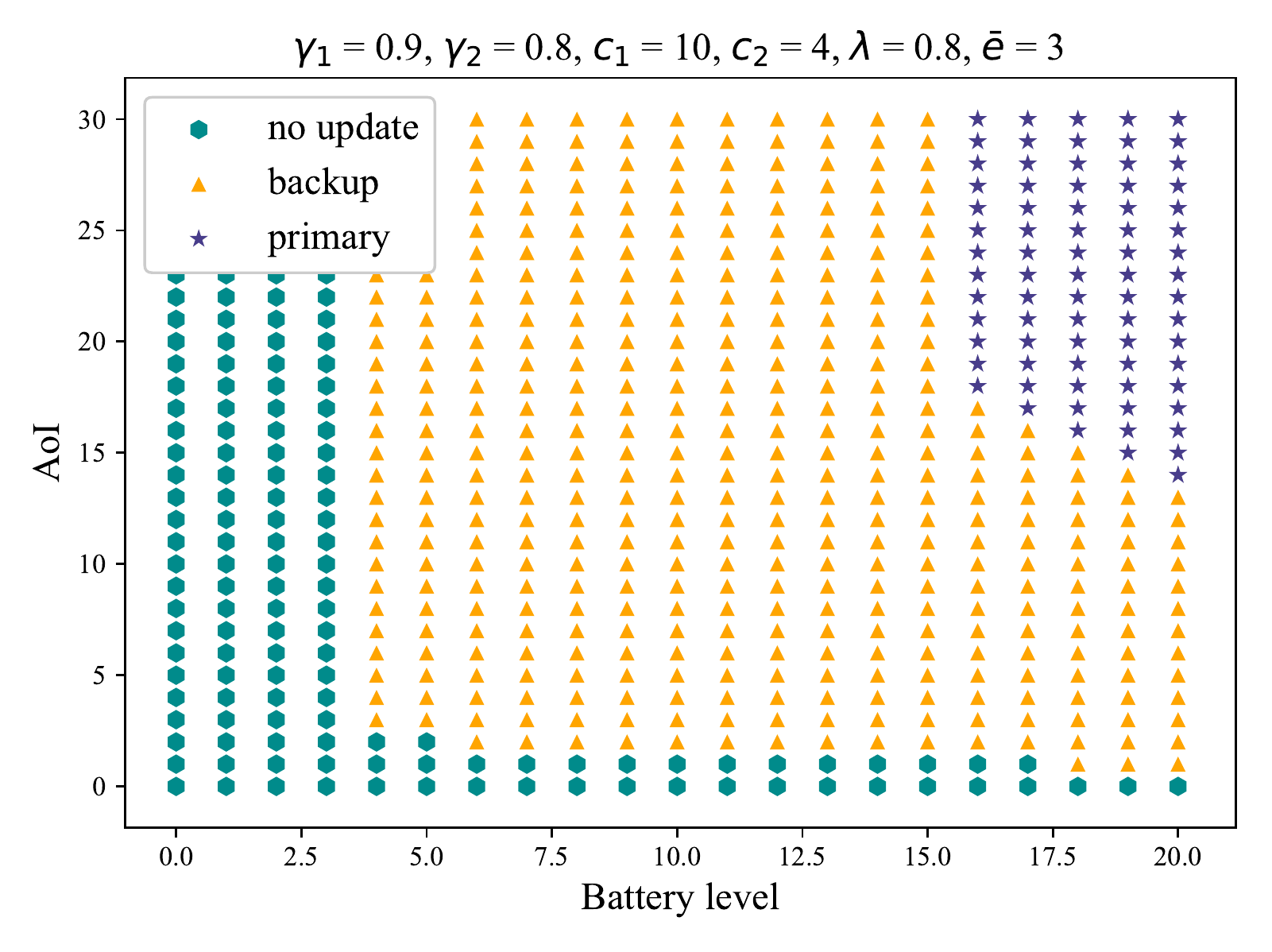}
		\label{fig:l08_g08_cr_04}%
	}
	\subfigure[Cost ratio = 0.8, $\lambda = 0.8$, $\gamma_2 = 0.8$]{%
		\includegraphics[width=0.14\textwidth]{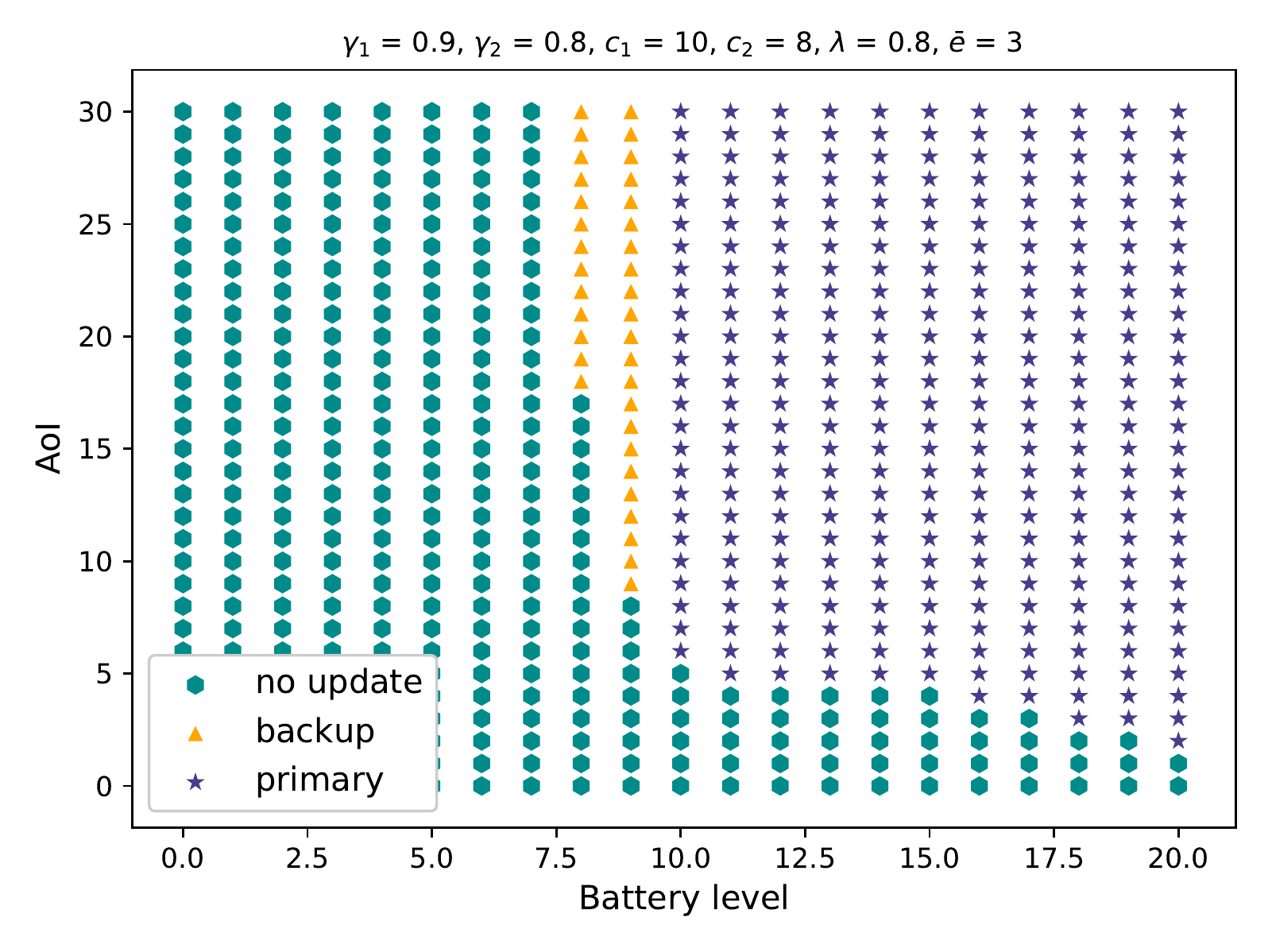}
		\label{fig:l08_g08_cr_08}%
	}
	\caption{Illustration of the optimal policy for different energy cost ratios $c_2/c_1$.}
	\label{fig:map_cost_ratio}
\end{figure}

The relative value of an information source can be measured by the portion of the states, in which the monitoring node chooses to exploit this source. To demonstrate this, we vary the cost ratio among the sources, $c_2/c_1$, and study the optimal policy obtained through RVI.  We see in Fig. \ref{fig:map_cost_ratio} that, when the cost ratio increases, the number of states at which the backup source is utilized shrinks, and the monitoring node opts to remain idle in most of the states. The disappearance of the backup source from the optimal action set is more rapid, if it is characterized by low reliability, $\gamma_2$ (see Figs. \ref{fig:l02_g02_cr_00} - \ref{fig:l02_g02_cr_08}, \ref{fig:l08_g02_cr_00} - \ref{fig:l08_g02_cr_08}). 

\begin{figure}[!t]	
	\vspace{-0.2cm}
	\centering
	\includegraphics[width=0.98\columnwidth]{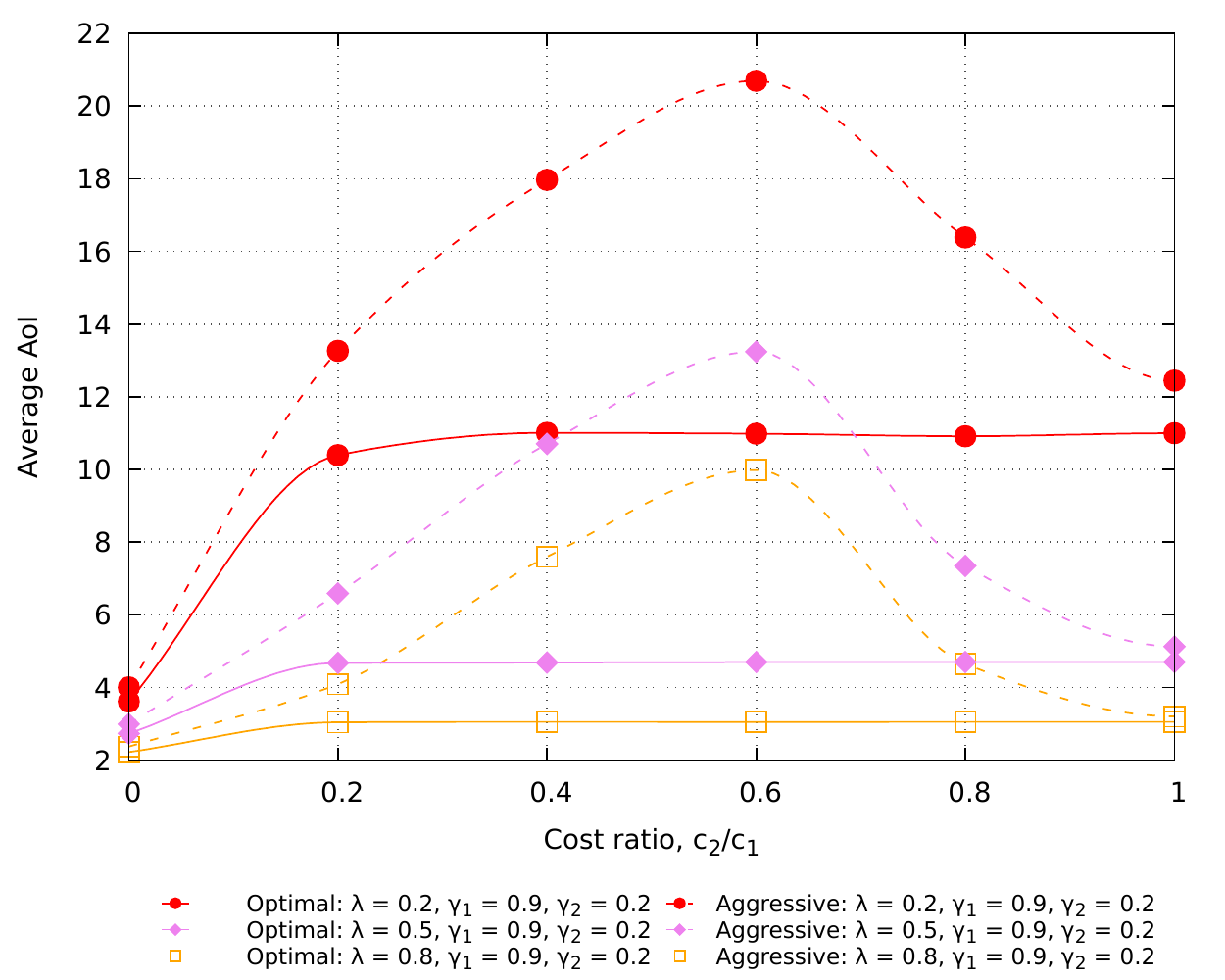}
	\caption{Dependency of average AoI and energy cost ratio for $\gamma_2 = 0.2$.}
	\label{fig:cost_vs_aoi_02}	
\end{figure}

\begin{figure}[!t]
	\vspace{-0.3cm}	
	\centering
	\includegraphics[width=0.98\columnwidth]{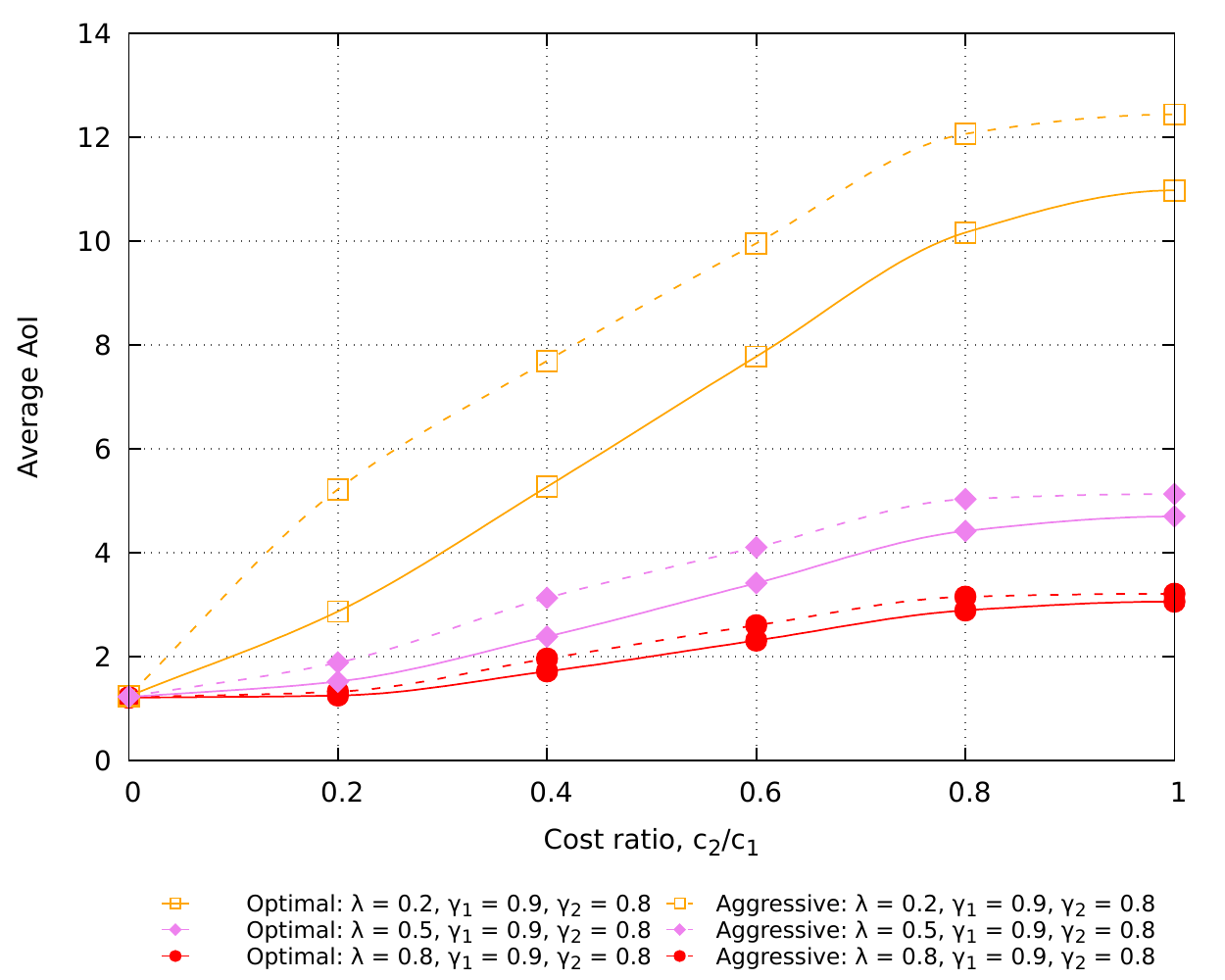}
	\caption{Dependency of average AoI and energy cost ratio for $\gamma_2 = 0.8$.}
	\label{fig:cost_vs_aoi_08}	
\end{figure}

\begin{figure}[!t]
	\vspace{-0.3cm}	
	\centering
	\includegraphics[width=0.98\columnwidth]{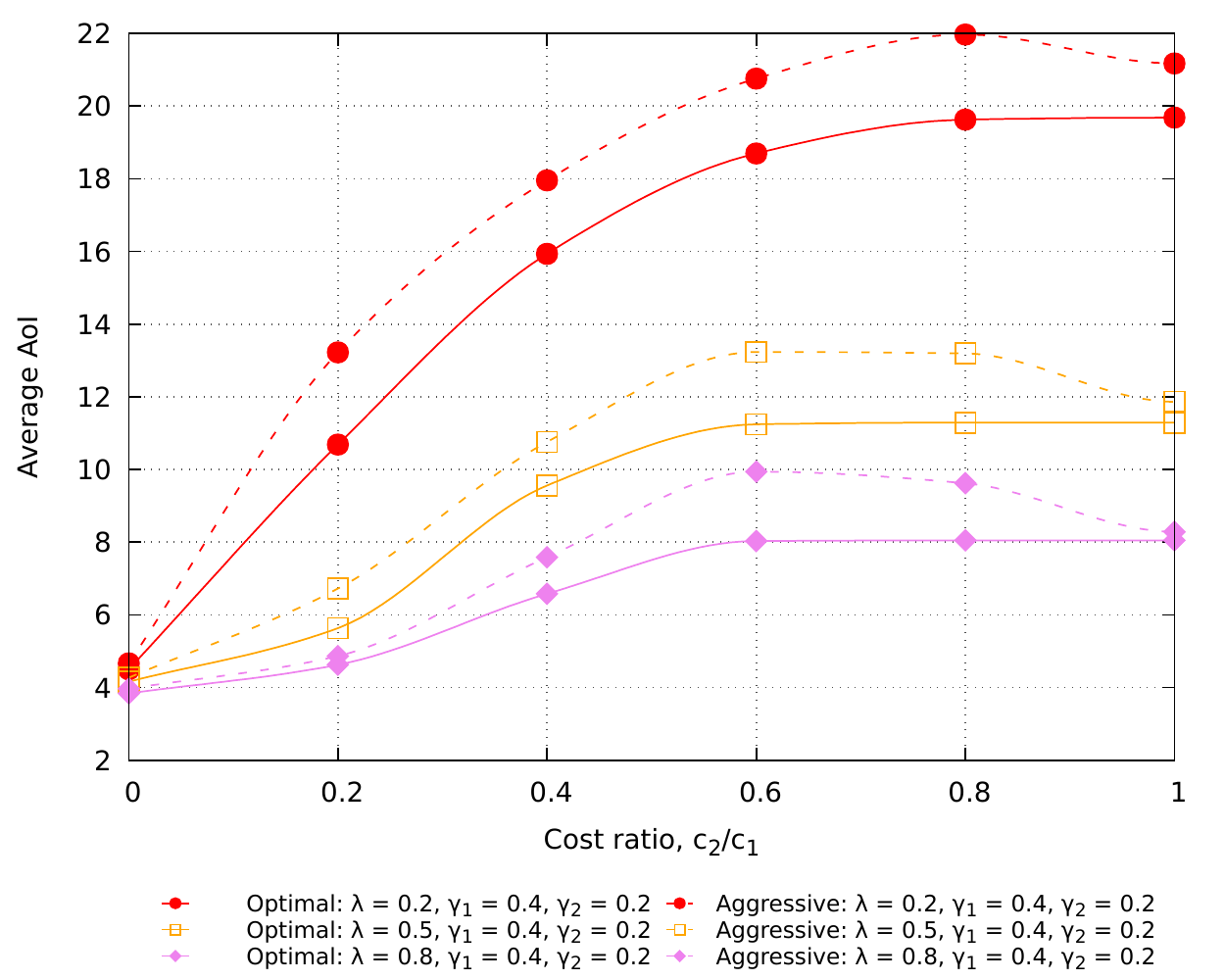}
	\caption{Dependency of average AoI and energy cost ratio for $\gamma_1 = 0.4$.}
	\label{fig:cost_vs_aoi_04}	
\end{figure}

The relation between the average AoI and cost ratio is shown in Figs. \ref{fig:cost_vs_aoi_02} - \ref{fig:cost_vs_aoi_08}. Predictably, the optimal average AoI grows when the cost ratio increases, but it saturates at a certain value, beyond which the backup source is not utilized at all. On the other hand, the average AoI increases quite rapidly at low values of the cost ratio. Moreover, for low values of $\lambda$, i.e., low energy generation rate, the saturation of the optimal average AoI happens at lower values of the cost ratio (see Fig. \ref{fig:cost_vs_aoi_02}). At lower values of source reliability, $\gamma_2$, the average AoI achieved by the aggressive policy does not have an intuitive behavior (see Fig. \ref{fig:cost_vs_aoi_02}). Up to a certain point (when $c_1 - c_2 {>} \bar{e}$), increasing usage of the backup source causes the average AoI to grow. After some point (when $c_1 - c_2 {\leq} \bar{e}$), the system starts to be more energy conserving, i.e., starts to reserve energy for getting updates from the more reliable primary source, and the average AoI starts decreasing. If we set  $\gamma_1 = 0.4$, a similar behaviour in average AoI is observed; however, the average AoI saturates at higher values of the cost ratio compared to the case when $\gamma_1 = 0.9$ (Fig. \ref{fig:cost_vs_aoi_02}).

\subsection{Energy harvesting}

\begin{figure}[!t]%
	\subfigure[$\lambda = 0.2$, cost ratio = 0.8, $\gamma_2 = 0.2$]{%
		\includegraphics[width=0.14\textwidth]{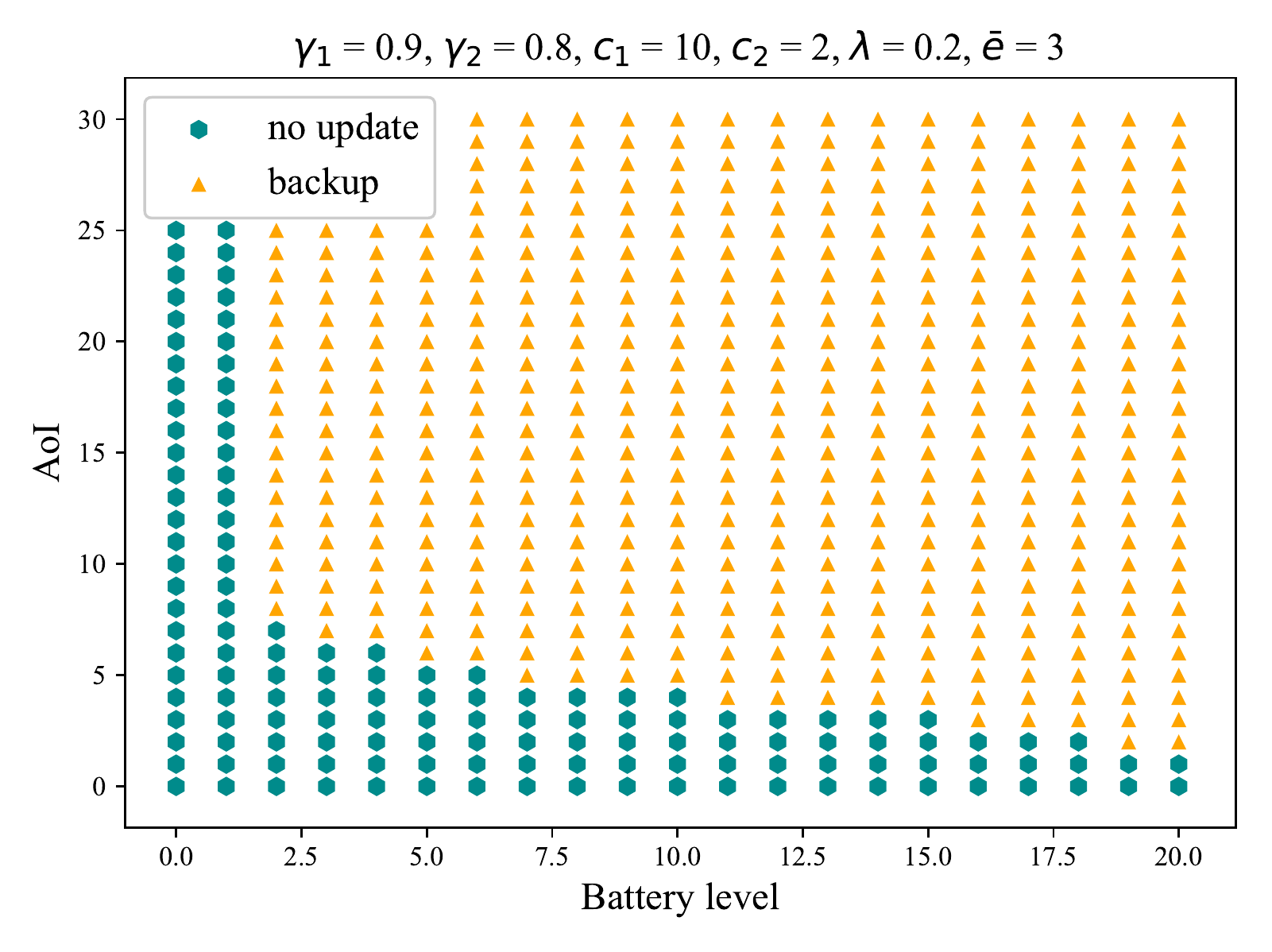}
		\label{fig:cr02_g08_l_02}%
	}
	\hspace*{\fill}
	\subfigure[$\lambda = 0.4$, cost ratio = 0.8, $\gamma_2 = 0.2$]{%
		\includegraphics[width=0.14\textwidth]{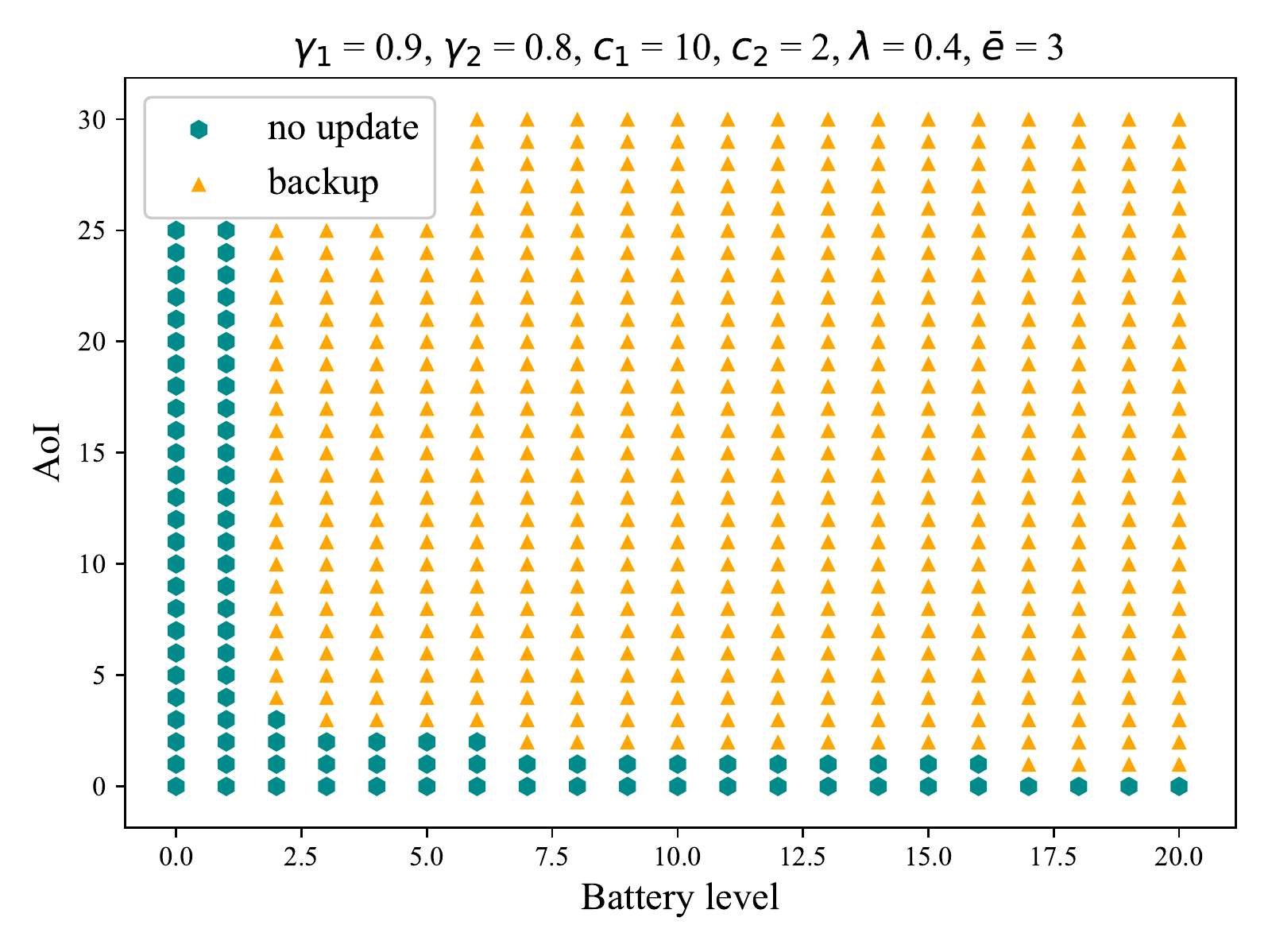}
		\label{fig:cr02_g08_l_04}%
	}
	\subfigure[$\lambda = 0.8$, cost ratio = 0.8, $\gamma_2 = 0.2$]{%
		\includegraphics[width=0.14\textwidth]{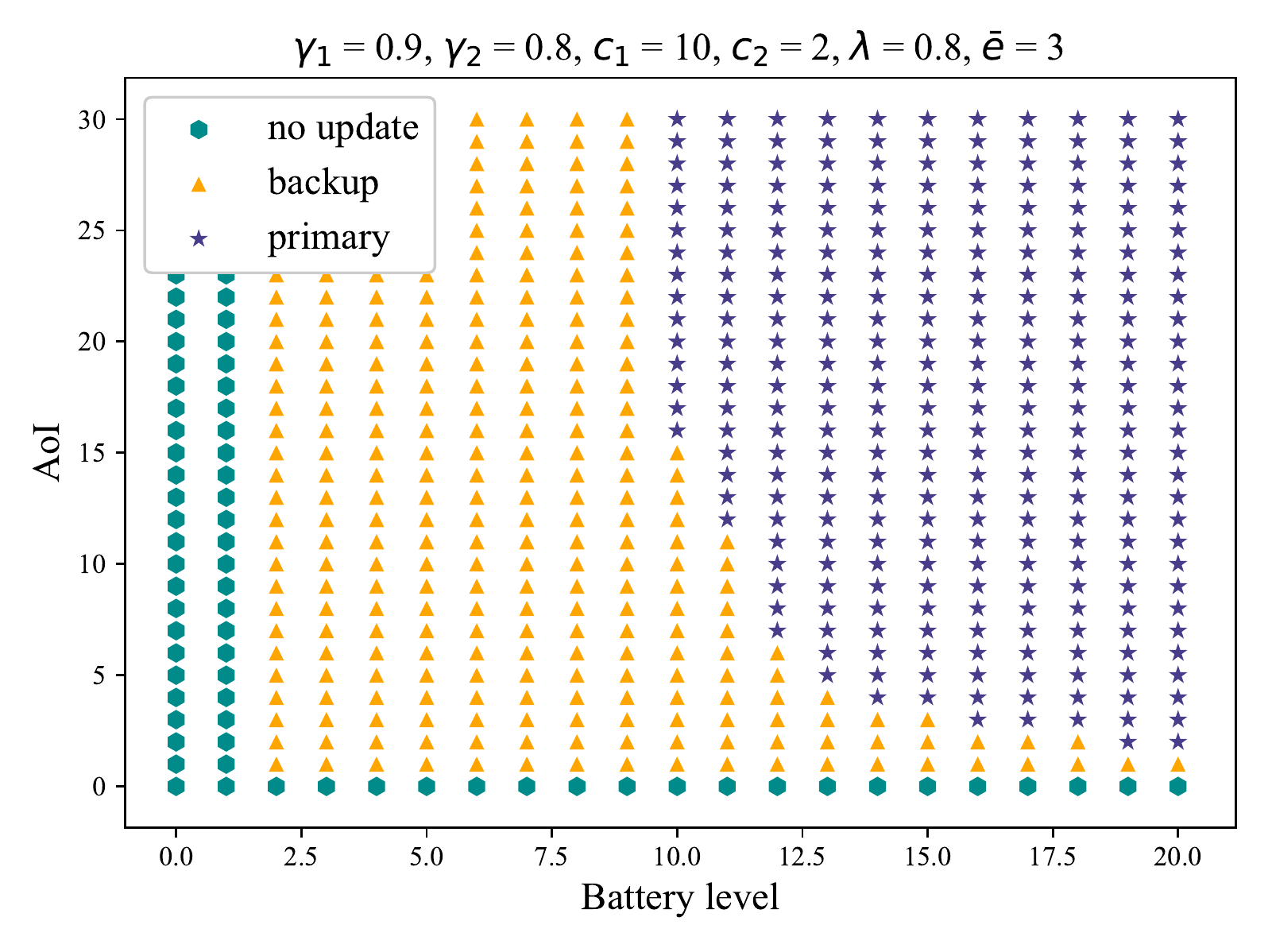}
		\label{fig:cr02_g08_l_08}%
	}
		\subfigure[$\lambda = 0.2$, cost ratio = 0.4, $\gamma_2 = 0.2$]{%
		\includegraphics[width=0.14\textwidth]{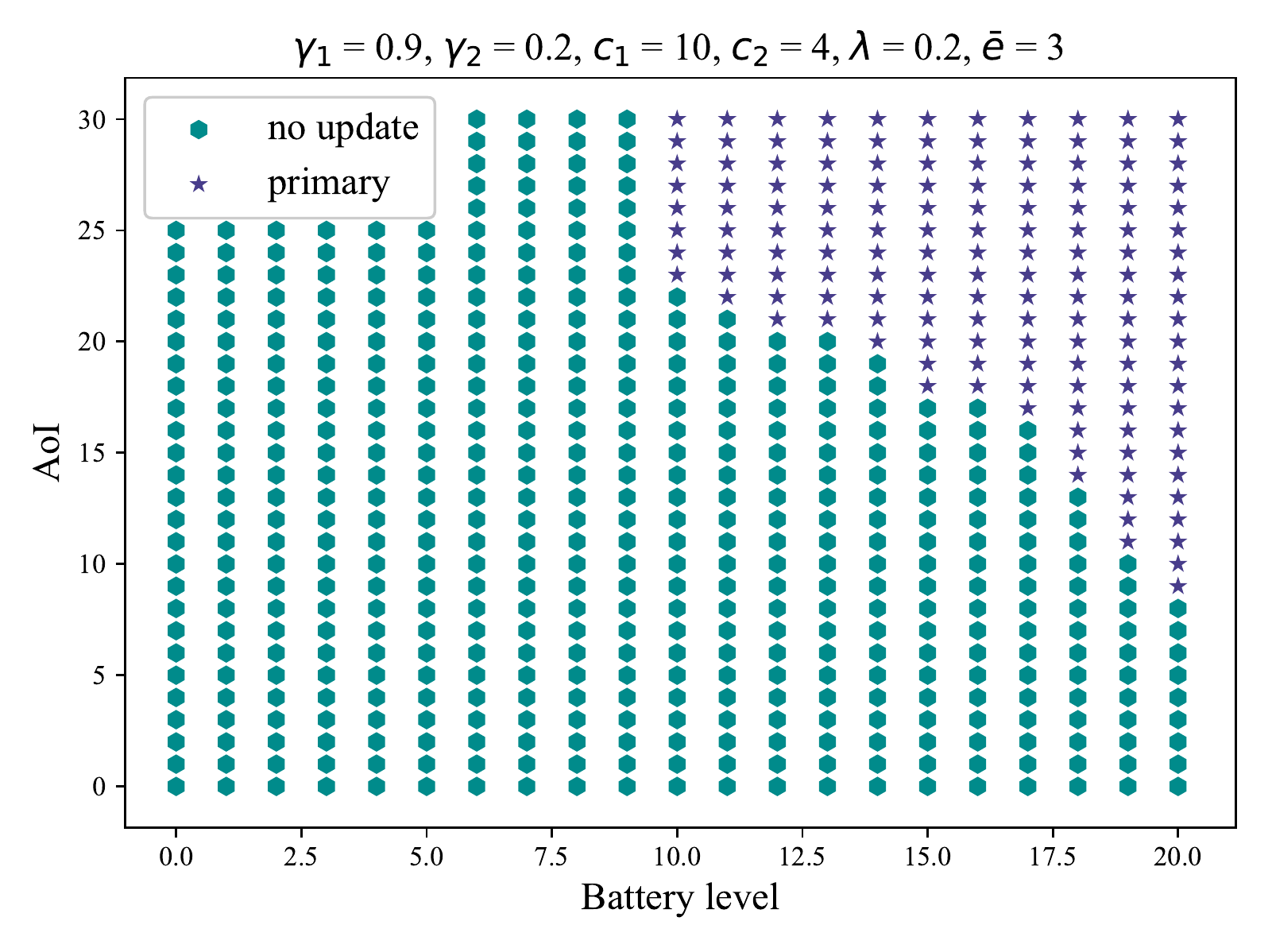}
		\label{fig:cr04_g02_l_02}%
	}
	\hspace*{\fill}
	\subfigure[$\lambda = 0.4$, cost ratio = 0.4, $\gamma_2 = 0.2$]{%
		\includegraphics[width=0.14\textwidth]{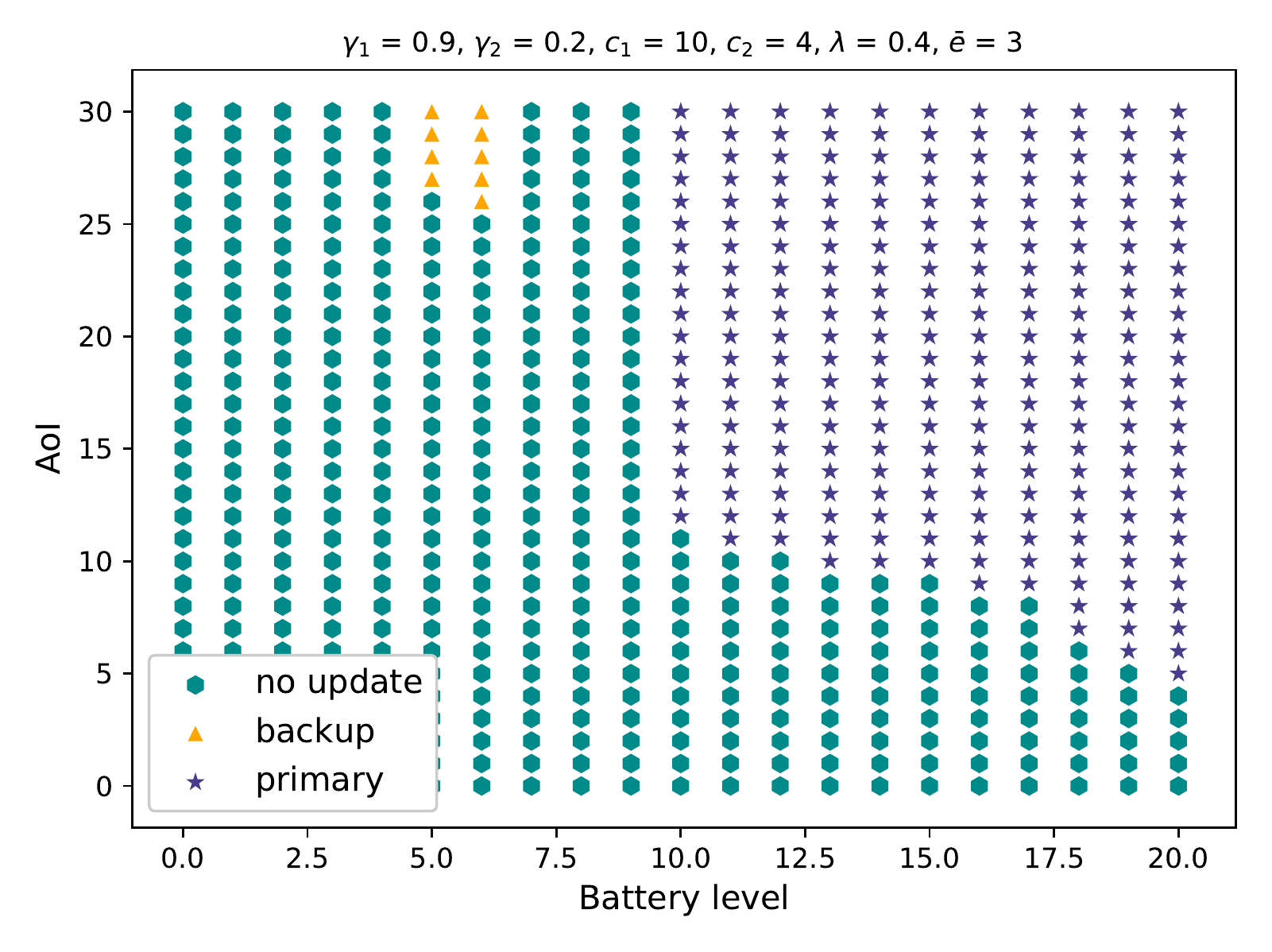}
		\label{fig:cr04_g02_l_04}%
	}
	\subfigure[$\lambda = 0.8$, cost ratio = 0.4, $\gamma_2 = 0.2$]{%
		\includegraphics[width=0.14\textwidth]{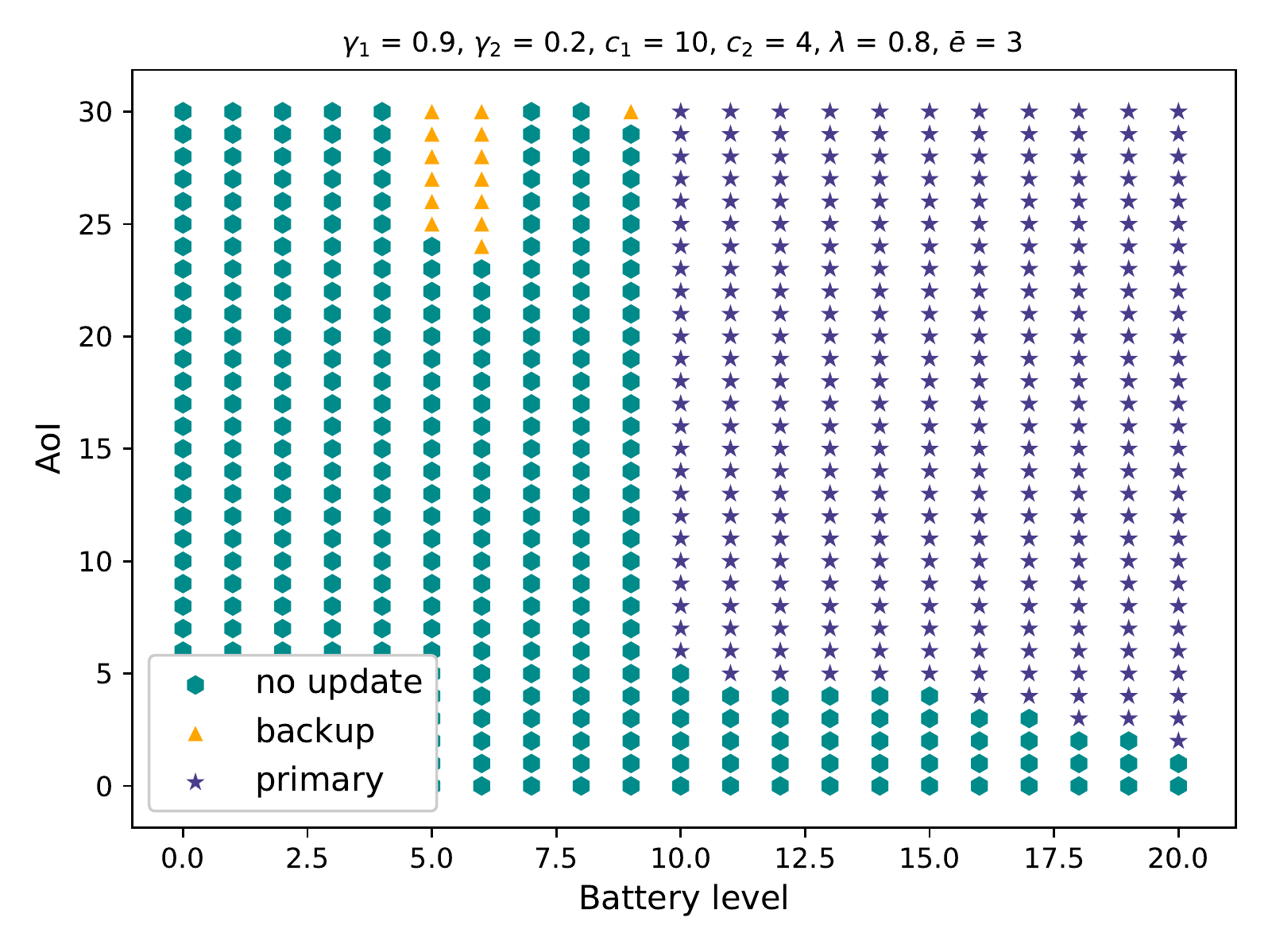}
		\label{fig:cr04_g02_l_08}%
	}
	\caption{Illustration of the optimal policy for different energy harvesting rates $\lambda$.}
	\label{fig:map_harvesting}
	\vspace{-0.4cm}
\end{figure}

Another important parameter that impacts the optimal solution is the energy harvesting rate, $\lambda$. With increasing $\lambda$ the monitoring node has tendency to request an update rather than staying idle (see Fig. \ref{fig:map_harvesting}). Furthermore, increasing energy harvesting capabilities enables the monitoring node to request updates more often from the primary information source, and reduces the utility of the backup source, which gradually disappears from the optimal solution. 

Some system configurations are characterized by having a `pocket' region, e.g., see Fig. \ref{fig:cr04_g02_l_04} and \ref{fig:cr04_g02_l_08}. This situation is observed when the reliability of the backup source is quite low and the energy harvesting rate is sufficiently high. In this case, the energy buffer can recover in a short amount of time, which enables the monitoring node to request an update from a primary source, instead of an extremely unreliable backup source. 

\begin{figure}[!t]
	\centering
	\includegraphics[width=0.98\columnwidth]{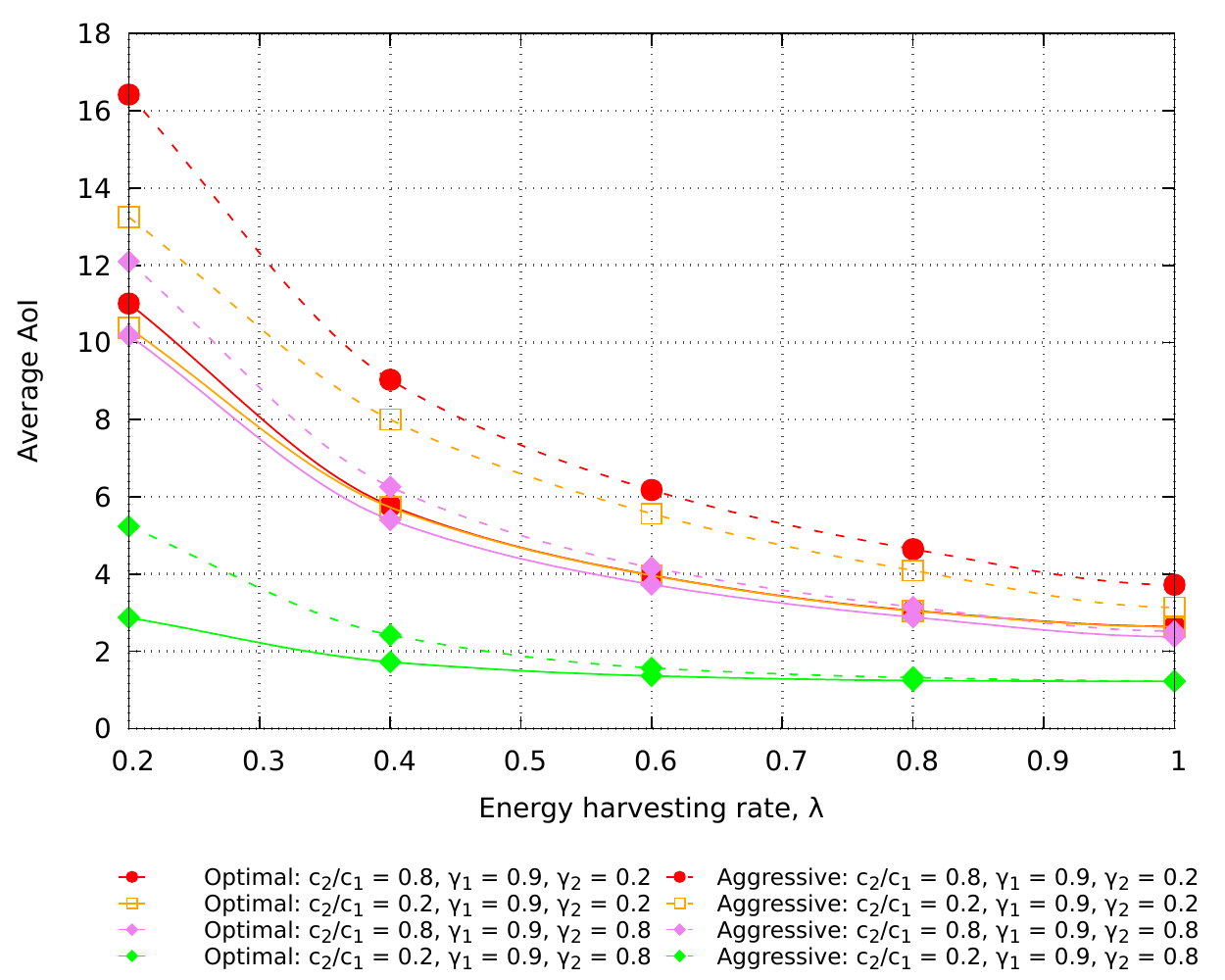}
	\caption{Average AoI as a function of the energy harvesting rate, $\lambda$.}
	\label{fig:harv_vs_aoi}
	\vspace{-0.4cm}	
\end{figure}

The dependence of the average AoI on $\lambda$ is demonstrated in Fig. \ref{fig:harv_vs_aoi}. As expected, the increase in the energy harvesting rate leads to a decrease in the achievable AoI.

\subsection{Reliability of information sources}

In Fig. \ref{fig:map_cost_ratio} we can also observe the evolution of the optimal solution as the reliability of the backup source, $\gamma_2$, increases. The increase in $\gamma_2$ leads to an increase in the number of states in which the backup source is queried. In other words, the utility of the backup source also increases.

\begin{figure}[!t]
	\centering
	\includegraphics[width=0.95\columnwidth]{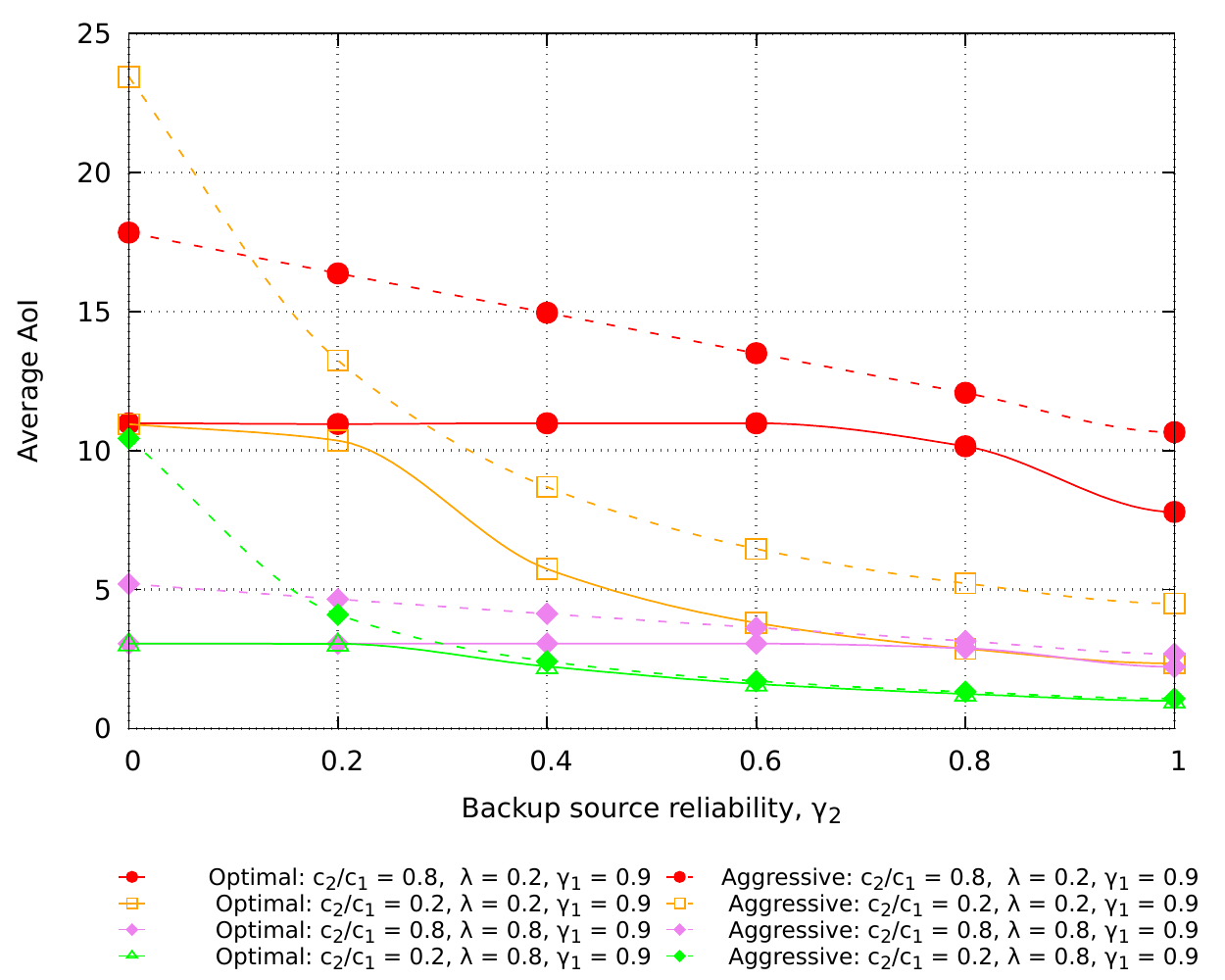}
	\caption{Average AoI as a function of the backup source reliability, $\gamma_2$.}
	\label{fig:gamma_vs_aoi}
	\vspace{-0.5cm}	
\end{figure}

The dependence of average AoI on $\gamma_2$ is shown in Fig. \ref{fig:gamma_vs_aoi}. As expected, the increase in the energy harvesting rate decreases the achievable average AoI. However, if $c_2/c_1$ is high, then the increase in $\gamma_2$ does not severely affect the average AoI. As the backup source has a high cost, then the primary source prevails in the optimal solution, and the reliability of the backup source does not affect the average AoI significantly. If both the cost ratio and the energy harvesting rate are low, then the backup source becomes more preferable as its reliability increases. Therefore, in this case we observe a significant drop in average AoI (see Fig. \ref{fig:gamma_vs_aoi}).   

\subsection{Efficiency evaluation}

\begin{figure}[!t]%
	\subfigure[$\lambda = 1.0$, cost ratio = 0.2, $\gamma_2 = 1.0$]{%
		\includegraphics[width=0.215\textwidth]{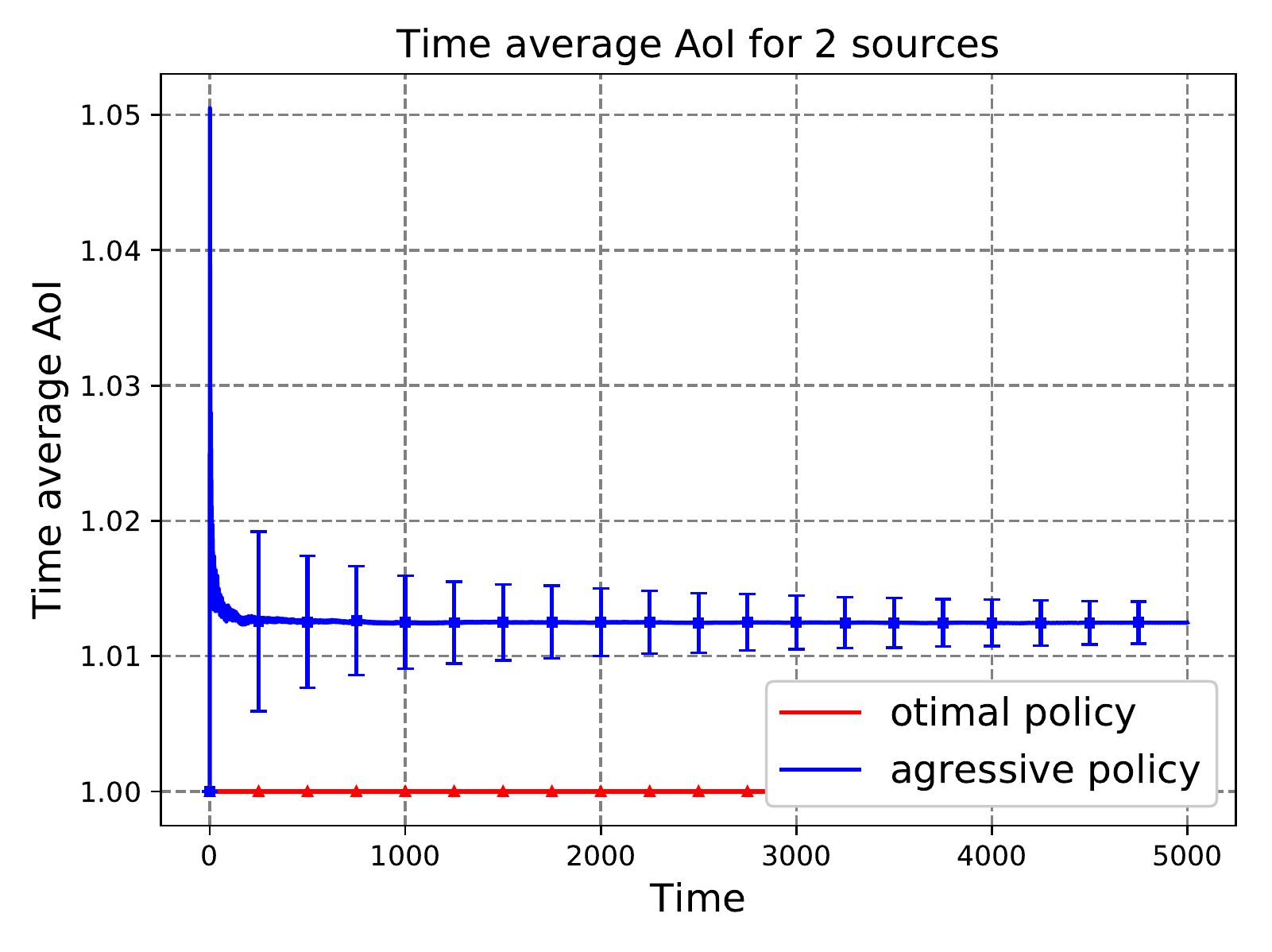}
		\label{fig:conv_1}%
	}
	\hspace*{\fill}
	\subfigure[$\lambda = 0.2$, cost ratio = 0.8, $\gamma_2 = 0.2$]{%
		\includegraphics[width=0.215\textwidth]{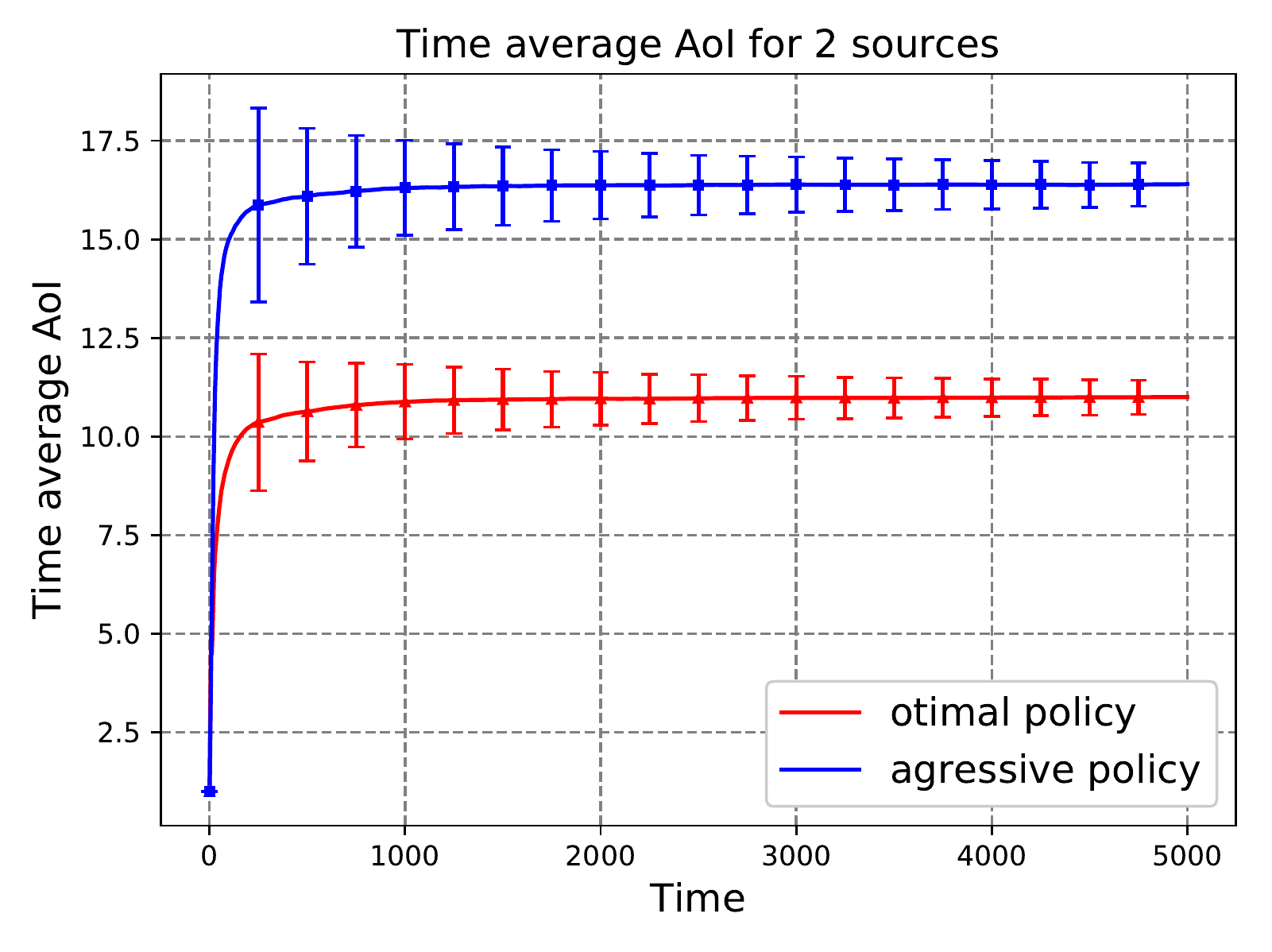}
		\label{fig:conv_2}%
	}
	\caption{Average AoI vs. time for the aggressive and optimal policies.}
	\label{fig:convergence}
	\vspace{-0.5cm}
\end{figure}

Finally, in Fig. \ref{fig:convergence} we compare the performance of the optimal and aggressive policies in terms of the average AoI. The convergence time for both policies are similar, and does not exceed 200 time slots. 

We observe that the gap between the average AoI achieved by the aggressive and optimal policies gets higher as the energy harvesting rate increases (Fig. \ref{fig:harv_vs_aoi}), i.e., if the energy arrivals to the system are relatively stable, then the aggressive policy can be as effective as the optimal one. Similarly, there is no gain in average AoI if $c_2/c_1 = 0$. $c_2/c_1$ does not significantly influence the relative performance of the optimal policy over the aggressive one, since the gap remains relatively constant as a function of $c_2/c_1$ (Fig. \ref{fig:cost_vs_aoi_02}-\ref{fig:cost_vs_aoi_08}). Generally speaking, since the backup source is less expensive but also less reliable than the primary one, the optimal policy tends to
preserve energy when convenient in order to use the primary source,
while the aggressive policy would always use the backup source
whenever possible. Thus, the gap between the two policies shrinks as
the backup source improves its reliability. However, if $c_2/c_1$
increases, the gap remains larger.

\subsection{Discussion}
\vspace{-0.1cm}	
We observe that the structure of the optimal solution varies depending on the characteristics of the environment and system parameters. In particular, we consider the energy harvesting rate as an environmental characteristic; the reliability of the information sources and the associated costs as system parameters. Two types of solution structure (behavior) can be distinguished: \emph{pocket region}, or appearance of the buffer (or accumulating) region in the optimal action set, where the monitoring node chooses to stay idle in order to gain extra energy, and \emph{monotonic disappearance} of a source from the optimal action set. 

Results reported above answer the question when employing a backup source is beneficial in reducing average AoI. Low values of cost ratio, as well as high reliability of the backup source are key requirements to be met in order to integrate a backup source to the system.  Improving environmental characteristics can reduce the need for the backup source, but also the benefits from employing the optimal policy. Sometimes improving the environmental characteristics (for instance, device relocation) can be a solution, instead of increasing the complexity of the system by adding extra backup devices. 

\vspace{-0.3cm}
\section{Conclusions}
\label{sec:con}
\vspace{-0.1cm}	
We have investigated a monitoring node that can query two distinct sources of information, a primary and a backup source, to receive status updates  of an underlying process of interest. We formulated this problem as an MDP, and derived the optimal policy that minimizes the average AoI. We compared the performance of the optimal policy with that of the aggressive policy, which tries to query the most expensive source it can afford, and demonstrated that the gain from the optimal policy increases as the energy harvesting rate decreases or the backup source characteristics become worse (i.e., decreasing reliability or increasing cost). 

We have also shown that employing an alternative source of information is justified  when the energy cost of requesting from the backup source is relatively low and its reliability is high. 

As a future work, the case with more than two sources of information will be considered with more than two possible AoI states that can be received from each of the sources.

\vspace{-0.3cm}
\bibliographystyle{IEEEtran}

\end{document}